\documentclass[aps,pra,twocolumn,reprint,floatfix, superscriptaddress]{revtex4-2}

\usepackage{subfigure}
\usepackage{psfrag,graphicx}
\usepackage{pict2e}
\usepackage{dcolumn}
\usepackage{amsmath,amssymb,braket}
\usepackage{bm}
\usepackage{color}
\usepackage{latexsym}
\usepackage{epstopdf}
\usepackage{color}
\usepackage[english]{babel}
\UseRawInputEncoding
\usepackage{amsfonts}
\usepackage{bm}
\usepackage{natbib}
\usepackage{bbold}

\usepackage{enumerate}
\definecolor{myblue}{rgb}{0.2,0.2,0.8}
\definecolor{myzard}{cmyk}{0,0,0.05,0}
\definecolor{mywhite}{rgb}{1,1,1}
\definecolor{mywhite}{rgb}{1,1,1}
\definecolor{myred}{rgb}{1,0.,0.3}
\usepackage[colorlinks=true,citecolor=myblue,linkcolor=myred]{hyperref}

\definecolor{mygrey}{gray}{0.35}
\definecolor{myblue}{rgb}{0.2,0.2,0.8}
\definecolor{myzard}{cmyk}{0,0,0.05,0}
\definecolor{mywhite}{rgb}{1,1,1}
\definecolor{mywhite}{rgb}{1,1,1}
\definecolor{myred}{rgb}{1,0.,0.3}

\def\be{\begin{equation}}
\def\ee{\end{equation}}
\def\ba{\begin{align}}
\def\enda{\end{align}}
\def\bi{\begin{itemize}}
\def\ei{\end{itemize}}

\def\beq{\begin{equation}}
\def\beq{\begin{equation}}
\def\eeq{\end{equation}}

\def\cc{{\rm c}}
\def\dd{{\rm d}}

\def\pp{{\textbf p}}
\def\nn{{\textbf n}}

\def\kk{{\textbf k}}
\def\qq{{\textbf q}}
\def\cc{{\textbf c}}
\def\dd{{\textbf d}}
\def\rr{{\textbf r}}

\begin{document}

\title{Quantum electrodynamics in  anisotropic and tilted Dirac photonic lattices}

\author{J.~Redondo-Yuste}
\affiliation{Institute of Fundamental Physics IFF-CSIC, Calle Serrano 113b, 28006 Madrid, Spain.}

\author{Mar\'ia Blanco de Paz}
\affiliation{Donostia International Physics Center, 20018 Donostia-San Sebastián, Spain}

\author{P.~A. Huidobro}
\affiliation{Instituto de Telecomunicações, Insituto Superior Tecnico-University of Lisbon, Avenida Rovisco Pais 1,  1049-001 Lisboa, Portugal.}

\author{A.~Gonz\'alez-Tudela}
\email{a.gonzalez.tudela@csic.es}
\affiliation{Institute of Fundamental Physics IFF-CSIC, Calle Serrano 113b, 28006 Madrid, Spain.}

\begin{abstract}

One of the most striking predictions of quantum electrodynamics is that vacuum fluctuations of the electromagnetic field can lead to spontaneous emission of atoms as well as photon-mediated interactions among them. Since these processes strongly depend on the nature of the photonic bath, a current burgeoning field is the study of their modification in the presence of photons with non-trivial energy dispersions, e.g., the ones confined in photonic crystals. A remarkable example is the case of isotropic Dirac-photons, which has been recently shown to lead to non-exponential spontaneous emission as well as dissipation-less long-range emitter interactions. In this work, we show how to further tune these processes by considering anisotropic Dirac cone dispersions, which include tilted, semi-Dirac, and the recently discovered type II and III Dirac points. In particular, we show how by changing the anisotropy of the lattice one can change both the spatial shape of the interactions as well as its coherent/incoherent nature. Finally, we discuss a possible implementation where these energy dispersions can be engineered and interfaced with quantum emitters based on subwavelength atomic arrays. 
\end{abstract}

\maketitle

\section{Introduction}

The extraordinary electronic properties of graphene are directly related to the emergence of isotropic Dirac-cone energy dispersions in the electronic band-structure of these systems~\cite{castroneto09a}. These properties have sparkled the interest in finding other materials where these energy dispersions appear~\cite{Wehling2014,Wang2015}, as well as in finding ways of modifying them to tune the emergent behaviour. For example, by introducing tunneling anisotropies, the Dirac cones can be displaced in reciprocal space, and even be annihilated by making two of them overlap~\cite{Hasegawa2006,Zhu2007,Volovik2007,Dietl2008,Wunsch2008,Montambaux2009,Pereira2009,DeGail2012,Lim2012,Delplace2013,rechtsman13a,Dutreix2013,Adroguer2016,Montambaux2018}. This overlap generates a new type of energy band touching linear in one direction and quadratic in the other, labeled as semi-Dirac point, which leads to unusual magnetic field dependence of electron transport~\cite{Dietl2008} or anisotropic polariton transport~\cite{Real2020}. Another example is the case of tilted Dirac dispersions that has been predicted to appear in quinoid and hydrogenated graphene~\cite{Goerbig2008,Lu2016,Katayama2006,Hirata2016}, and which can be harnessed for valley filtering~\cite{Nguyen2018} or generating photocurrents~\cite{Chan2017}. Besides, when the tilting reaches a critical value, its associated Fermi surface changes from being a point to a line becoming what have been labeled as Type II and III Dirac points~\cite{Soluyanov2015,Xu2015a,Deng2016,Huang2016,Noh2017,Huang2018,Liu2017b,Perrot2018,Milicevic2019a,Mizoguchi2020,Jin2020,Kim2020a}, which are believed to be instrumental to enhance the superconducting gap~\cite{Li2017}, to probe flat-band physics~\cite{Leykam2018} and quantum chaos~\cite{Chen2020}, or even study analogue black holes in solid-state environments~\cite{Huang2018,Volovik2016}.

Triggered by these exciting properties, the interest in engineering Dirac energy dispersions has expanded to fields beyond the electronic realm. One paradigmatic example is the case of photonics, where such Dirac energy dispersions have been proposed and engineered to obtain unconventional photon transport properties~\cite{Real2020,haldane08a,sepkhanov07a,zandbergen10a,zhang08a,huang11a,bravo12a,Yang2018,Hu2018,Milicevic2019a,Kim2020a}, among other phenomena. One of the latest frontiers of the field is to consider the interplay of these structured photons with quantum emitters, since they are expected to modify substantially quantum optical phenomena such as spontaneous emission or photon-mediated interactions. These changes have already been explored in two recent works for isotropic Dirac photons~\cite{Gonzalez-Tudela2018,Perczel2020a}. There, it was shown how a single emitter tuned to the Dirac point can display a non-exponential relaxation in spite of the vanishing density of states at this point~\cite{Gonzalez-Tudela2018}. Besides, this unconventional relaxation was linked to the emergence of a power-law localized photonic mode around the emitter, labeled as quasi-bound state~\cite{Pereira2006,Wehling2007,Dutreix2013}, which can ultimately mediate power-law interactions between emitters with no associated dissipation~\cite{Gonzalez-Tudela2018,Perczel2020a}. These long-range interactions are especially attractive because they can be harnessed for quantum information and simulation purposes~\cite{vodola14a,gong16a,eldredge17a,chen19a,Tran2020c}. Since more complex Dirac-energy dispersions have started being designed and built within the photonic context~\cite{Yang2018,Hu2018,Milicevic2019a,Kim2020a}, a timely question is to understand their impact in these quantum optical phenomena when quantum emitters interact with such photonic lattices.

In this work, we target precisely this problem by considering a system where quantum emitters interact with a honeycomb photonic lattice with a controlled degree of anisotropy and tilting of the Dirac points. For the case with anisotropic Dirac points we observe how indeed the spatial shape of the photon-mediated interactions can be modified without losing neither its long-range nor its decoherence-free character, as it occurs with the case of 3D Weyl points~\cite{Garcia-Elcano2020,Garcia-Elcano2021,ying19a}. Something similar occurs for the tilted case, until a critical value of the tilting when the density of states becomes finite and the resonant nodal lines lead to directional radiation patterns. The latter will translate into anisotropic collective dissipative terms when many emitters couple to the bath, as the ones found in photonic Van-Hove singularities~\cite{Gonzalez-Tudela2017b,Gonzalez-Tudela2017a,galve17a}, that will lead to strong super/subradiant effects~\cite{dicke54a}. We illustrate all these effects using a minimal tight-binding model of the photonic bath where these energy dispersions appear, that allows to extract numerical and analytical insight of these phenomena. However, we also finally provide a possible way of probing these effects using a combination of subwavelength atomic arrays~\cite{Rui2020,Glicenstein2020} and additional impurity atoms~\cite{Masson2020,Patti2021,Brechtelsbauer2020}.

The manuscript is structured as follows: in Sections~\ref{sec:system} and~\ref{sec:theoryframework} we describe the system under study and the theoretical framework that we will use to characterize it, respectively. In Section~\ref{sec:review}, we review the results known for isotropic Dirac dispersions, and then in Sections~\ref{sec:anisotropy} and~\ref{sec:tilted} we analyze the change of the photon-mediated interactions with the degree of anisotropy and tilting of the bath energy dispersion. Then, in Section~\ref{sec:implementation}, we analyze the band-structure of several subwavelength atomic arrays configurations where some of these energy dispersions appear. Finally, we conclude and summarize our findings in Section~\ref{sec:conclu}.

\section{System~\label{sec:system}}

\begin{figure}[!tb]
  \centering
  \includegraphics[width=0.45\textwidth]{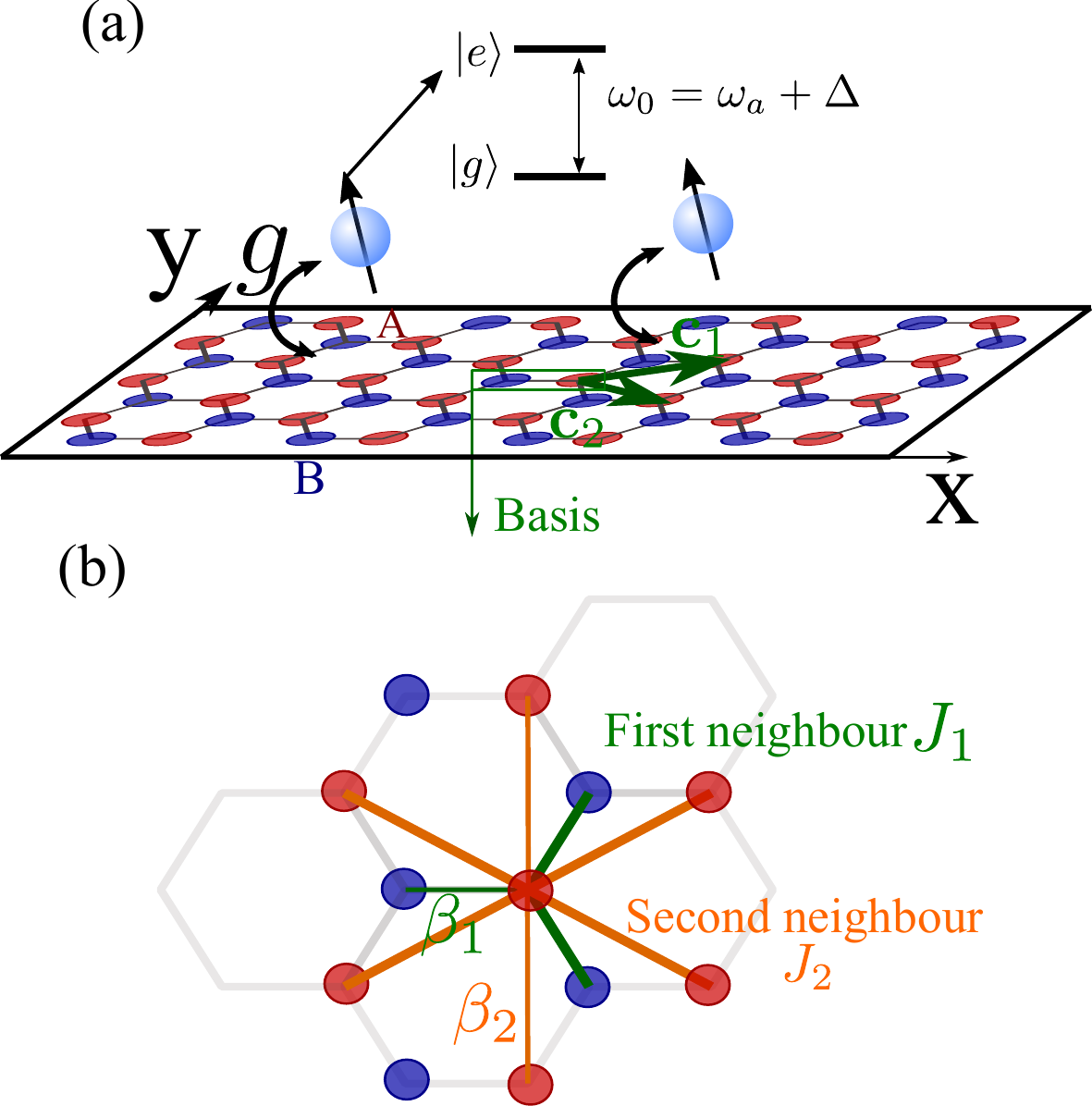}
  \caption{a) Scheme of the system: one or many two-level emitters are coupled locally with strength $g$ to a photonic lattice described by a bipartite lattice with primitive vectors $\cc_{1/2}$ (b) Hopping structure of the bipartite bath: we consider nearest (green) and next-to-nearest (orange) hoppings with overall strength $J_{1,2}$, respectively. The strength of the nearest (and next-nearest neighbour) hoppings along one direction can be modified through an anisotropy parameter  $\beta_1$ $(\beta_2)$, respectively.}
  \label{fig:scheme}
\end{figure}

The system that we consider along this manuscript (see scheme of Fig.~\ref{fig:scheme}(a)) consists of a collection of $N_e$ quantum emitters that, for simplicity, we consider as two-level systems with a ground, $g$, and an excited state $e$ with an energy difference $\hbar \omega_0$. The corresponding Hamiltonian then reads (assuming $\hbar\equiv 1$ for the rest of the manuscript):
\begin{align}
    H_S=\omega_0 \sum_{j}\sigma_{ee}^j\,,\label{eq:HS}
\end{align}
where we use the notation $\sigma_{\alpha\beta}^j=\ket{\alpha}_j\bra{\beta}_j$ for the operator acting on the $j$-th emitter. Through their optical transition, these emitters couple to a photonic bath site, described by a bosonic annihilation (creation) operator $O^{(\dagger)}_\nn$, through the standard light-matter Hamiltonian:
\begin{align}
    H_I=\sum_j \left(g_j O_{\nn_j} \sigma_{eg}^j+\mathrm{H.c.}\right)\,,\label{eq:HII}
\end{align}
where $g_j$ denotes the strength of the coupling for the $j$-th emitter, and $\nn_j$ denotes the position of the bath where the emitter couples to. For simplicity we will assume this coupling to be emitter-independent, $g_j\equiv g$. Note also that in Eq.~\eqref{eq:HII} we are implicitly assuming that the light-matter coupling is i) local, since the emitter can only couple a single bosonic mode at a given position; and ii) excitation-conserving, since we are neglecting the counter-rotating terms ($O_{\nn_j} \sigma_{ge}^j$,\dots). Both approximations are generally a good description in the optical regime~\cite{breuer-petruccione}, or even an exact one in the case of simulated light-matter couplings with matter-waves~\cite{devega08a,navarretebenlloch11a,krinner18a}, that is one of the most promising implementations to observe these effects.

The non-trivial part of the problem stems from the photonic bath. To be able to extract analytical and numerical insight of the physics, it is convenient to adopt a discrete photonic lattice model where the bath is described as an array of coupled cavities. Since we want to describe Dirac-like dispersions the array is disposed in a honeycomb geometry as depicted in Fig.~\ref{fig:scheme}(a). This array can be described by a bipartite lattice with bosonic operators $a_\nn,b_\nn$ for the A/B sublattices, respectively. We label the positions $\nn=(n_1,n_2)$ with $n_i\in \mathbb{Z}$ in terms of the primitive lattice vector displacements, i.e., $\nn\rightarrow \sum_{i=1}^2 n_i \cc_i$, with $\cc_i$ being the primitive vectors of the lattice $\cc_{1/2}=3/2\hat{e}_x\pm \sqrt{3}/2\hat{e}_y$ (taking the lattice constant as the unit of length). Imposing periodic boundary conditions, we can define $O_\kk=1/\sqrt{N}\sum_\nn e^{-i\kk\cdot\nn}O_\nn$, for $O=a,b$, being $N$ the total number of sites of each sublattice, and $\kk\rightarrow \sum_{i=1}^2 k_i \dd_i$ with $\dd_i$ being the primitive vectors of the reciprocal lattice, i.e, $\cc_i\cdot\dd_j=\delta_{ij}$, with $k_i\in [-\pi,\pi)$ in $2\pi/N$ steps. With these operators the bath Hamiltonian can always be written as:
\begin{align}
    H_B=\sum_\kk (a_\kk^\dagger\, b_\kk^\dagger)\tilde{H}_B(\kk)\begin{pmatrix}
a_\kk \\
b_\kk
\end{pmatrix}\,\,
\end{align}
where $\tilde{H}_B(\kk)$ is a $2\times 2$ matrix that can be expanded in terms of the Pauli matrices: $\tilde{H}_B(\kk)=(\omega_a+h_I(\kk))\mathbf{1}+\sum_{\alpha=x,y,z}h_\alpha(\kk)\sigma_\alpha$. Here, the functions $h_\alpha(\kk)$ contains the information about the hopping structure of the bath, and $\omega_a$ is the energy of cavities describing the lattice. In what follows, we will set this energy as the energy reference of the problem, i.e., $\omega_a\equiv 0$, and replace $\omega_0\rightarrow \Delta=\omega_0-\omega_a$ in Eq.~\eqref{eq:HS}, which will now denote the detuning of the optical transition of the emitters with respect to the bath energy modes.
This model can be easily diagonalized as follows:
\begin{equation}
  H_B=\sum_\kk \left(\omega_u(\kk) u^\dagger_\kk u_\kk+\omega_l(\kk) l^\dagger_\kk l_\kk\right)\,,~\label{eq:HI}
\end{equation}
yielding two bands with energies $\omega_{u/l}(\kk)=h_I(\kk)\pm \sqrt{|h_x(\kk)|^2+|h_y(\kk)|^2+|h_z(\kk)|^2}=h_I(\kk)\pm\omega(\kk)$. The relation between the original operators $a_\kk,b_\kk$ and the eigen-operators $u_\kk,l_\kk$ is obtained through the unitary matrix that diagonalizes $\tilde{H}_B(\kk)$, such that%
\begin{align}
    \begin{pmatrix}
a_\kk \\
b_\kk
\end{pmatrix}=\tilde{U}_\kk \begin{pmatrix}
u_\kk \\
l_\kk
\end{pmatrix}=\begin{pmatrix}
\cos\left(\frac{\theta_\kk}{2}\right) e^{-i\phi_\kk} & \sin\left(\frac{\theta_\kk}{2}\right)e^{-i\phi_\kk} \\
\sin\left(\frac{\theta_\kk}{2}\right) & -\cos\left(\frac{\theta_\kk}{2}\right) 
\end{pmatrix}\begin{pmatrix}
u_\kk \\
l_\kk
\end{pmatrix}
\end{align}
where the angles $\theta_\kk,\phi_\kk$ are given by:
\begin{align}
    \cos\left(\theta_\kk\right)&=\frac{h_z(\kk)}{\omega(\kk)}\,,\\
    \tan\left(\phi_\kk\right)&=\frac{h_y(\kk)}{h_x(\kk)}\,.
\end{align}

Once this diagonalization is done, typically it is more convenient to write $H_I$ of Eq.~\ref{eq:HI} in terms of the eigen-operators $u_\kk,l_\kk$:
\begin{align}
    H_I=\sum_j \left( B_{\nn_j} \sigma_{eg}^j+\mathrm{H.c.}\right)\,,
\end{align}
with
\begin{align}
  B_{\nn_j}= \frac{g}{\sqrt{N}}\sum_\kk
    e^{i (\kk\cdot \nn_j-\phi_\kk)}\left[\cos(\theta_\kk/2)u_\kk +\sin(\theta_\kk/2)l_\kk\right]\,,
\end{align}
if the $j$-th emitter is coupled to the A-sublattice, i.e., $\nn_j\in S_A$, whereas:
\begin{align}
     B_{\nn_j}=\frac{g}{\sqrt{N}}\sum_\kk
    e^{i \kk\cdot \nn_j}\left[\sin(\theta_\kk/2)u_\kk - \cos(\theta_\kk/2)l_\kk\right]
\end{align}
if $\nn_j\in S_B$.

From all these previous expressions, we see that the information of both the emergent band-structure, and how the emitter couples to the bath is hidden in the $h_{\alpha}(\kk)$, which depend on the particular hopping structure of the bath. Since we want to explore both anisotropic and tilted Dirac cones we will consider the following hopping structure (see Fig.~\ref{fig:scheme}(b)):
\begin{itemize}
    \item We consider nearest neighbour hoppings (in green in the figure) of overall strength $-J_1$, with the possibility of a weakest one with strength $-\beta_1 J_1$ for the one corresponding within the same unit cell.
    
    \item We also consider the possibility of having next-to-nearest neighbour hoppings with overall strength $-J_2$, with the possibility of having a weaker link, $-\beta_2 J_2$ for the hoppings.
\end{itemize}

This hopping choice leads to the following $h_\alpha(\kk)$:
\begin{align}
    h_I(\kk)&=-2 J_2\left[\cos(k_1)+\cos(k_2)+\beta_2 \cos(k_1-k_2)\right]\,,\label{eq:hIk}\\
    h_z(\kk)&=0\,,\\
    h_x(\kk)&=-J_1\left[\beta_1+\cos(k_1)+\cos(k_2)\right]\,,\\
    h_y(\kk)&=-J_1\left[\sin(k_1)+\sin(k_2)\right]\,.
\end{align}

Notice that since $h_z(\kk)=0\rightarrow \theta_\kk=\pi/2$, which simplifies the calculations.

\section{Theoretical framework~\label{sec:theoryframework}}

Along this work we are interested in studying how individual and collective spontaneous emission are modified by the interaction with the photons supported by such anisotropic Dirac photonic lattices. Since the total Hamiltonian describing the global system, i.e., $H=H_S+H_B+H_I$, conserves the number of excitations, the dynamics and eigenstates of the system can be characterized independently in each excitation manifold. For example, in the single excitation subspace, the following ansatz:
\begin{align}
    \ket{\Psi}=\Big[\sum_j C_{e,j}\sigma^j_{eg}+\sum_{O=a,b}\sum_\pp C_{O,\pp}O^\dagger_{\pp}\Big]\ket{\mathrm{vac}}\,,\label{eq:ansatz}
\end{align}
describes both the shape of all possible eigenstates, as well as the dynamics of the system when initialized with a single excitation. In this work, we will use this ansatz for the case when a single emitter is coupled to the bath for two purposes:
\begin{itemize}
    \item To find the emergent qubit-photon bound-states of the system~\cite{bykov75a,john90a,kurizki90a}. These can be obtained by finding the solutions of  $H\ket{\Psi_{\mathrm{BS}}}=E_\mathrm{BS}\ket{\Psi_{\mathrm{BS}}}$, with $E_\mathrm{BS}\neq \omega_{u/l}(\kk)$. Using this ansatz one can find that the energy of these bound states is given by:
    \begin{align}
        E_\mathrm{BS}=\Delta+\Sigma_e^\alpha(E_\mathrm{BS})\,,\label{eq:solBS}
    \end{align}
    where $\Sigma^{A/B}_{e}(z)$ is self-energy of the emitter, which for bipartite bath reads:
\begin{align}
    \Sigma^{A/B}_e(z)=\frac{g^2}{N}\sum_\kk\left[\frac{z-h_I(\kk)}{(z-h_I(\kk))^2-\omega(\kk)^2}\right]\,.\label{eq:sigmae}
\end{align}

Here, the superindex denotes the sublattice to which the emitter is coupled to. However, since the bath has chiral symmetry, $h_z(\kk)=0$, the single emitter self-energy is the same no matter to which sublattice the emitter is coupled to, i.e.,  $\Sigma^{A/B}_e(z)\equiv \Sigma_e(z)$. An analytical expression for the wavefunction coefficients $C_{O,\pp}$ can also be obtained. For example, if the emitter couples to the $A$ sublattice, the photonic component of the bound state reads (up to a normalization coefficient):
\begin{align}
   C^{A}_{a,\pp}&= \frac{1}{N}\sum_\kk \left[\frac{E_\mathrm{BS}-h_I(\kk)}{(E_\mathrm{BS}-h_I(\kk))^2-\omega(\kk)^2}\right]e^{i\kk\cdot\pp}\,,\label{eq:cAA}
 \end{align}
 
 \begin{align}\label{eq:cAB}
C^{A}_{b,\pp}&= \frac{1}{N}\sum_\kk \frac{\omega(\kk)}{(E_\mathrm{BS}-h_I(\kk))^2-\omega(\kk)^2}e^{i(\kk\cdot\pp-\phi_\kk)}\,,
\end{align}
where the super-index denotes the sublattice that the emitter is coupled to, and the sub-index the mode to which the spatial coefficient belongs in Eq.~\eqref{eq:ansatz}. The spatial distribution of these wavefunctions is very relevant because it eventually dictates the shape of the emitter-emitter interaction when many emitters couple to the bath. In particular, it can be shown that when the emitters' energy lie within a vanishing density of states region and the photons can be adiabatically eliminated (assuming Born-Markov conditions are satisfied~\cite{breuer-petruccione}), the effective emitters' dynamics induced by the photons is given by~\cite{douglas15a,Gonzalez-Tudela2015b,Bello2019a}:
\begin{equation}
    \label{eq:effH}
    H_\mathrm{eff}\approx \sum_{ij}G_{ji}^{\alpha\beta} \sigma_{eg}^j\sigma_{ge}^i\,,
\end{equation}
where $G_{ij}^{\alpha\beta}$, which represents the interaction between the $i$-th emitter coupled to the $\alpha$-sublattice and the $j$-th emitter coupled to the $\beta$-sublattice, is given (up to a constant) by $G^{AA/BB}_{ij}\propto C^{A/B}_{a/b,\rr_i-\rr_j}$ and $G^{AB}_{ij}\propto C^{A}_{b,\rr_i-\rr_j}$. Thus, by studying already the bound state wavefunctions of a single emitter, one can already characterize the emergent coherent dipole interactions appearing when many emitters couple to the photonic bath.

\item We will also use the ansatz of Eq.~\eqref{eq:ansatz} to calculate the dynamics of the system after initializing the system in a given state, $\ket{\Psi(0)}$. For that one must solve the time-dependent Schr\"odinger equation:
    \begin{align}
        \ket{\Psi(t)}=e^{-i H t}\ket{\Psi(0)}\,,
    \end{align}
of the global system either by numerical or analytical means. For example, in the case of a single emitter initially excited, $\ket{\Psi(0)}=\sigma_{eg}\ket{\mathrm{vac}}$, the excited state population $C_e(t)$ can be expressed analytically as the following Laplace transform~\cite{CohenTannoudji1998}:
\begin{equation}
    C^\alpha_e(t)=-\frac{1}{2\pi i}\int_{-\infty}^\infty dE \frac{e^{-i E t}}{E+i 0^+-\Delta-\Sigma_e^{\alpha}(E+i 0^+)}\,,\label{eq:cet}
\end{equation}
depending on whether the emitter couples to the $\alpha=A,B$ sublattice. There, we observe how the single emitter self-energy not only determines the emergence of qubit-photon bound states, but also ultimately determines its dynamics. In fact, it is possible to recover the expected Markovian result by approximating: $\Sigma_e^{\alpha}(E+i 0^+)=\Sigma_e^{\alpha}(\Delta+i 0^+)=\delta\omega_e^\alpha(\Delta)-i\Gamma^\alpha_e(\Delta)/2$ in the integrand of Eq.~\eqref{eq:cet}, defining as $\delta\omega_e^\alpha, \Gamma^\alpha_e$ the renormalization of the emitter's energy and lifetimes due to the interaction with the bath. Within that approximation, the excited state population will decay exponentially to the ground state:
\begin{equation}
|C^\alpha_e(t)|^2\approx e^{-\Gamma_e(\Delta)t}\,,
\end{equation}
with $\Gamma_e(\Delta)$ being exactly the one predicted by Fermi's Golden Rule:
\begin{align}
    \Gamma_{e}^{\alpha}(\Delta)&=\pi g^2 \sum_{\kk,p=u/l}\delta(\Delta-\omega_p(\kk)) \,,
\end{align}
that is, proportional to $g^2$ and the density of states of the bath, $D(E)=\sum_{\kk,p=u/l} \delta(E-\omega_p(\kk))$. As the emitter decays to the ground state, it launches propagating excitations into the bath which can be characterized by studying $C_{p,\nn}(t)$. These dynamical spatial distributions can be also linked to the effective emitter's interactions induced by the photons. When the photons eventually propagate away from the emitter, as it is the case if their energies lie within the photonic band structure, the photon-induced interactions will eventually generate (individual and) collective dissipative terms, $\Gamma_{ij}$, that can lead to strong super/subradiant effects when many emitters couple to the bath~\cite{dicke54a}. The spatial shape of these terms, $\Gamma_{ij}$, is ultimately the same as the one of the propagating fields into the bath. This is why, again, by looking into the spontaneous emission of a single emitter, we will be able to extract information about the collective dissipative terms when many emitters couple to the bath.

\end{itemize}

\begin{figure}[tb]
        \centering
        
        \includegraphics[width=0.45\textwidth]{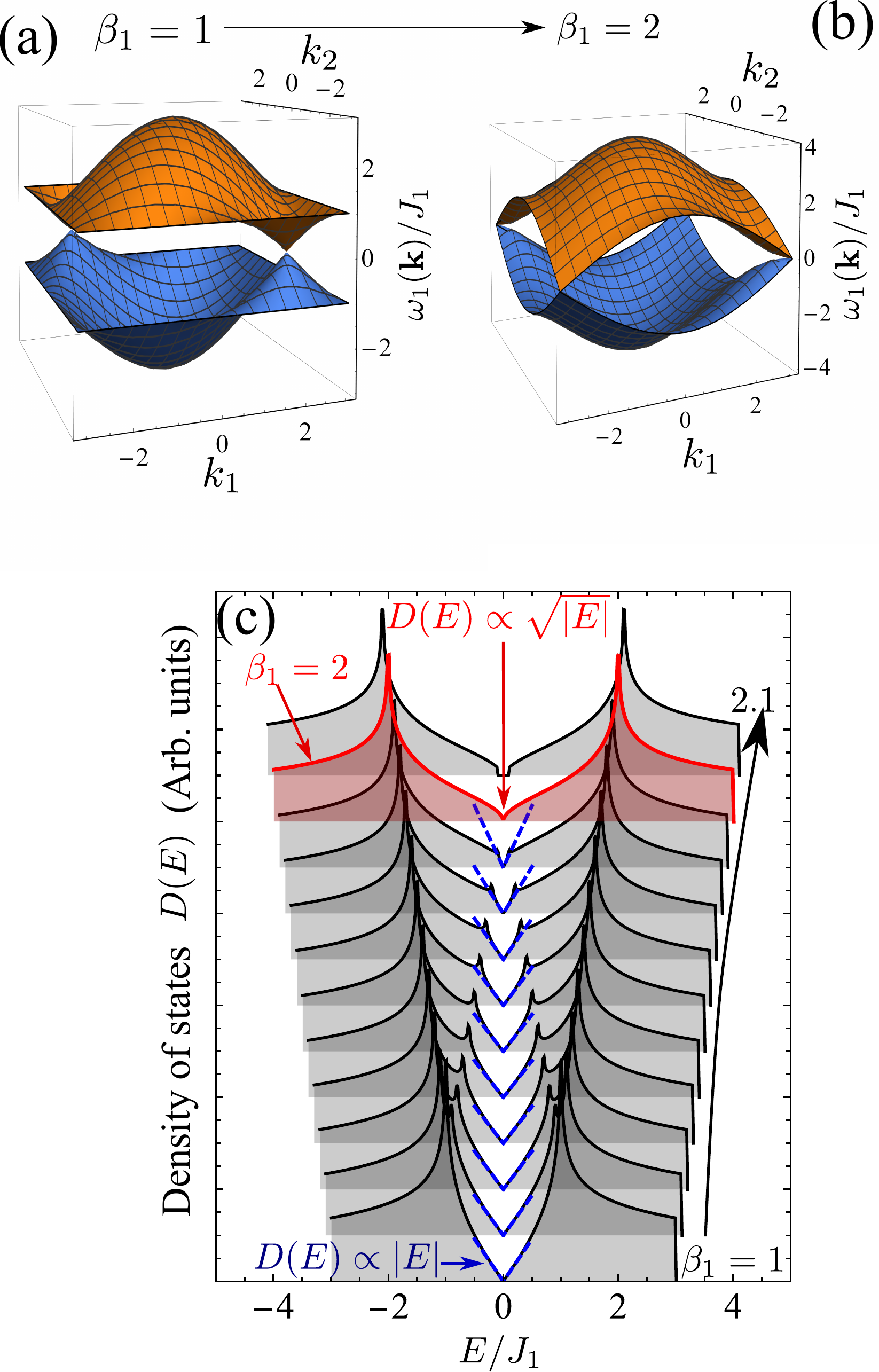}
        \caption{(a) Energy dispersion, $\pm \omega_\mathrm{iso}(\kk)$ for the isotropic nearest neighbour situation ($\beta_1=1$) where two linearly dispersive band-touching occurs, labeled as Dirac points. (b) Energy dispersion, $\pm\omega_1(\kk)$, for the anisotropic situation with $\beta_1=2$. The two Dirac points merge at the corners of the Brillouin zone forming a new band-touching, labeled as semi-Dirac point, that is parabolic in one-direction and linear in the other. (c) Evolution of the density of states of the bath, $D(E)$, with the anisotropy parameter $\beta_1$. The density of states maintains the linear dependence with the energy around the Dirac point, i.e., $D(E)\propto |E|$ (in dashed blue), until the critical value $\beta_1=2$ (in red) where it changes to $D(E)\propto \sqrt{|E|}$. For $\beta_1>2$ a finite gap opens around the Dirac energy.}
        \label{fig:figsemi1}
    \end{figure}
    
\section{Emitter dynamics and interactions in isotropic graphene~\label{sec:review}}

Before digging into more complex energy dispersions, let us summarize the main findings obtained for the situation of isotropic Dirac-cone dispersions~\cite{Gonzalez-Tudela2018,Perczel2020a}. These are obtained by setting $J_2=0, \beta_1=1$ in our model. This implies that $h_I(\kk)=0$, such that the spectrum is symmetric around the reference energy, i.e., $\omega_{u/k}(\kk)=\pm\omega_\mathrm{iso}(\kk)$, with:
\begin{align}
\omega_\mathrm{iso}(\kk)= J_1\sqrt{3+2\left[\cos(k_1)+\cos(k_2)+\cos(k_1-k_2)\right]}\,.
\end{align}

Note that for these coordinates, $(k_1,k_2)$, the two inequivalent Dirac points appear at $\mathbf{K},\mathbf{K'}=(2\pi/3)(\pm 1,\mp 1)$ (see Fig.~\ref{fig:figsemi1}(a)), and the energy dispersion around it is not isotropic $\omega_\mathrm{iso}(\mathbf{K}^{(')}+\qq)\approx J_1\sqrt{q_1^2+q_2^2-q_1 q_2}$. However, the isotropy can be immediately recovered by changing to the real space coordinates $(k_x=(k_1+k_2)/3,k_y=(k_1-k_2)/\sqrt{3})$ obtained by using that the reciprocal vectors in terms of real space axis read $\dd_{1/2}=\hat{e}_x/3\pm \hat{e}_y/\sqrt{3}$. In these coordinates, there are six Dirac points at $\mathbf{K}^{(')}$ (only two inequivalent ones) around which the energy dispersion is isotropic: $\omega_\mathrm{iso}(\mathbf{K}^{(')}+\qq)\approx \frac{3 J_1}{2}|\qq|$.

Focusing on the regime where the energy of the emitters is tuned to the Dirac point, i.e., $\Delta=0$, and using the theoretical framework explained in section~\ref{sec:theoryframework}, it was found that~\cite{Gonzalez-Tudela2018}:
\begin{itemize}

    \item Since $\Sigma_e^\alpha(0)=0$, then Eq.~\eqref{eq:solBS} has always a trivial solution at $E_\mathrm{BS}=0$. When looking at the spatial distribution of this bound-state solution, one can find that $C^{\alpha}_{\alpha,\pp}=0$, while:
\begin{align}
C^{A/B}_{b/a,\pp}&= \Big[e^{i\mathbf{K}\cdot\pp}I_\mathbf{K}(\pp)+e^{i\mathbf{K'}\cdot\pp}I_\mathbf{K'}(\pp)\Big]\label{eq:Gijcont}\,,
\end{align}
where $I_{\mathbf{K}^{(')}}(\pp)$ can be shown to scale in the limit of large distances as:
\begin{align}
\label{eq:Ik}
|I_{\mathbf{K}^{(')}}(\pp)|\propto \frac{1}{|\tilde{\pp}|}\,,
\end{align}
 with $\tilde{\pp}=\left(3(p_1+p_2)/2,\sqrt{3}(p_1-p_2)/2\right)$ (see Ref.~\cite{Gonzalez-Tudela2018}). This means that when the emitter couples to the A [B] sublattice, a localized mode emerges around it in the opposite sublattice with an isotropic power-law $1/r$-decay, plus some oscillatory factor that accounts for the interference of two integrals $I_{\mathbf{K},\mathbf{K}^{'}}$ in Eq.~\eqref{eq:Gijcont}, which does not modify the asymptotic scaling of the $|C^{A/B}_{a,b,\pp}|$. This very slow decay makes this bound-state very different from the standard bound-states found in other photonic band-gaps~\cite{bykov75a,john90a,kurizki90a,douglas15a,Gonzalez-Tudela2015b}, since its wavefunction is not integrable in the thermodynamic limit $N\rightarrow\infty$ (it diverges logarithmically). Since this bound state is similar to the one found in the context of graphene electronic transport~\cite{Pereira2006,Wehling2007,Dutreix2013} we will use the same nomenclature and label it as \emph{quasi-bound} state, to distinguish it from conventional qubit-photon bound states~\cite{bykov75a,john90a,kurizki90a,douglas15a,Gonzalez-Tudela2015b}. In what follows, we will see how this quasi-bound state will also lead to distinctive qualitative features of the individual and collective quantum emitter dynamics.
 
    \item The excited state probability amplitude of an initially single excited emitter coupled to the bath is approximately given by:
    \begin{equation}
    C_e(t)\approx R_0+C_\mathrm{NM}(t)\,,\label{eq:cetiso}
    \end{equation}
    where $R_0$ is the overlap of the initial state with the (quasi)-bound state, and $C_\mathrm{NM}(t)$ is a non-exponential contribution ($\propto 1/\log(t)$) appearing due to the non-analytical nature of the density of states around the Dirac point, i.e., $D(E)\propto |E|$. In standard photonic band-gaps~\cite{bykov75a,john90a,kurizki90a,douglas15a,Gonzalez-Tudela2015b}, $C_\mathrm{NM}(t)$ is typically hidden by the constant contribution $R_0$ which is generally very close to 1 and independent of system size. However, the non-integrable nature of the bound states appearing in Dirac photonic environments leads to a dramatically different behaviour. In particular, here $R_0$ depends on system size as follows~\cite{Gonzalez-Tudela2018}:
    \begin{align}
    R_0(N)=\frac{1}{1+\frac{g^2}{J_1^2}g(N)}\,,\label{eq:R0}
    \end{align}
    with $g(N)=\frac{J_1^2}{N}\sum_\kk\frac{1}{\omega_\mathrm{iso}(\kk)^2}\approx 0.2+0.37\ln(N)$. This means that in the thermodynamic limit $R_0(N\rightarrow \infty)\rightarrow 0$. Thus, in an infinite-size bath, one should observe only a purely non-Markovian relaxation $C_e(t)\approx C_\mathrm{NM}(t)$, instead of the no decay expected from the vanishing density of states at the emitter's frequency, i.e., $D(\Delta=0)=0$.
    
    \item Besides, when many emitters couple at these conditions to the bath, their associated quasi-bound state can mediate coherent long-range interactions, $G_{ij}^{\alpha\beta}\propto 1/|\rr_{ij}|$, as the ones predicted by Eq.~\eqref{eq:effH}. However, it was also shown that the finite overlap with the quasi-bound state, $R_0(N)$, ultimately renormalizes these interactions as follows:
    \begin{align}
    G_{ij}^{\alpha\beta}\propto   \frac{g^2 R_0(N)}{J_1 |\tilde{\rr}_{ij}|}\,.
    \end{align}
    
    Thus, for an infinite bath $G_{ij}^{\alpha\beta}\equiv 0 $.     Fortunately, since $R_0(N)$ goes to zero logarithmically with system size, i.e., $R_0(N)\propto 1/\ln(N)$, the quasi-bound state can effectively mediate interactions still in very large systems.
\end{itemize}

\begin{figure*}[!tb]
        \centering
        
        \includegraphics[width=\textwidth]{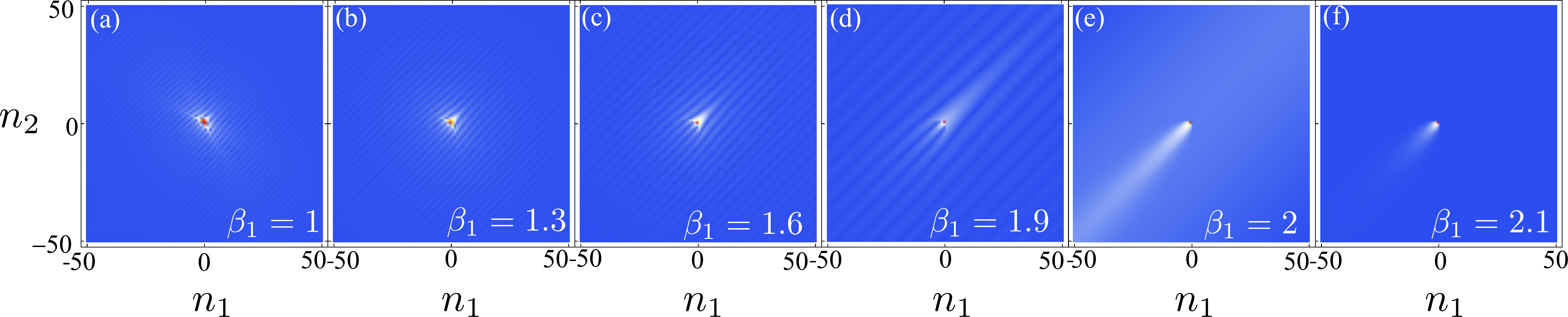}
        \caption{(a-f) Evolution of the quasi-bound state shape $|C^A_{b,\nn}|$ at $E_{\mathrm{BS}}=0$ for several values of the anisotropy parameters $\beta_1$, as depicted in the legend of each panel. Bath size for all panels is $N=500^2$.}
        \label{fig:figsemi2}
    \end{figure*}
   
In what follows, we will consider how these quasi-bound states, and their quantum optical consequences, get affected by the inclusion of anisotropies and tilting in the Dirac-energy dispersion

\section{Photon-mediated interactions in anisotropic and semi-Dirac points\label{sec:anisotropy}}

Let us first consider what happens when we include an anisotropy in the nearest-neighbour hoppings, e.g., by taking $\beta_1\neq 1$ in the hopping model of Fig.~\ref{fig:scheme}(b), and $J_2=0$. Note that here we are already choosing a particular direction of the anisotropy. However, the results can be extrapolated to the other choices just by considering a rotation of the real space coordinates.  In that case, it is still satisfied that $h_I(\kk)=h_z(\kk)=0$, such that one can use all the general expressions of the self-energies, bound-state wavefunctions, etc, derived for the isotropic case, but replacing $\omega_\mathrm{iso}(\kk)$ by the following anisotropic energy dispersion:
\begin{align}
    \frac{\omega_1(\kk)}{J_1}=\sqrt{2+\beta_1^2+2\cos(k_1-k_2)+2\beta_1 \left[\cos(k_1)+\cos(k_2)\right]}\,.
\end{align}

The main consequence of the anisotropy $\beta_1$ is that it provides a knob to move the position of the Dirac points in the Brillouin zone~~\cite{Hasegawa2006,Zhu2007,Volovik2007,Dietl2008,Wunsch2008,Montambaux2009,Pereira2009,DeGail2012,Lim2012,rechtsman13a,Adroguer2016,Montambaux2018}, $\mathbf{K}/\mathbf{K'}=(\pm \arccos(-\beta_1/2), \mp \arccos(-\beta_1/2))$, without altering its linear dispersion. This can be done until a critical anisotropy value, $\beta_1=2$, where the Dirac points merge forming a new type of band-crossing labeled as semi-Dirac point, linear in one direction, and parabolic in the other (see Fig.~\ref{fig:figsemi1}(b)).  This can be better understood making a rotation of the axis $q_{1/2}=(k_2\pm k_1)/\sqrt{2}$ and expanding the $H_B(\kk)$ around the $\mathbf{K},\mathbf{K'}$ to obtain:
\begin{align}
        \Tilde{H}_B(\mathbf{K}/\mathbf{K}^{'}+\mathbf{q})\approx v_{1}\sigma_y q_1\mp v_{2}\sigma_x q_2, \label{eq:expHbaniso1}
\end{align}
for $|\qq_i|\ll 1$. Here, $\sigma_{x,y}$ are Pauli matrices, and $v_{i}$ the effective velocities appearing in the different $q_i$-directions:
\begin{align}
        v_{2}=J_1\sqrt{2-\frac{\beta_1^2}{2}}, \quad v_{1}=J_1\frac{\beta_1}{\sqrt{2}}. \label{eq:v1v2}
    \end{align}

Like this, it is trivial to see that the energy dispersion maintains its linear dispersion around the Dirac point $\omega_1(\mathbf{K}^{(')}+\qq)\approx \sqrt{v_1^2 q_1^2+v_2^2 q_2^2}$, until $\beta_1=2$, where $v_2$ vanishes, such that the linear expansion of Eq.~\eqref{eq:expHbaniso1} is not valid anymore. 

This transition from a linear to a semi-Dirac band touching translates into important qualitative differences in magnitudes that govern the quantum optical phenomena of the system, such as the bath density of states. This is what we show in Fig.~\ref{fig:figsemi1}(c) where we plot $D(E)$ for increasing values of anisotropy ranging from $\beta_1=1$ (isotropic case) to $\beta_1=2.1$, including the semi-Dirac case, $\beta_1=2$ highlighted in red. There, we observe several important features:
\begin{itemize}
    \item For $\beta_1<2$, the density of states around the Dirac energy maintains both its singular nature ($D(0)=0$) and linear energy dependence, $D(E)\propto |E|$. In fact, using Eqs.~\eqref{eq:expHbaniso1}-\eqref{eq:v1v2} one can show that:
    \begin{align}
        D(E)\simeq \frac{|E|}{2\pi v_1v_2},
    \end{align}
    for $|E|\ll J_1$. This expression is the one we plot in dashed blue in Fig.~\ref{fig:figsemi1}(c), finding a very good agreement with the numerically obtained $D(E)$ for this energy range.
    
    \item Another important difference in the regime $1<\beta_1\le 2$ is the appearance of two additional Van-Hove singularities between the main ones of the isotropic case and the Dirac energy. These Van-Hove singularities move closer to the Dirac energy as $\beta_1\rightarrow 2$, until they collapse at the critical value where the two Dirac energies merge, i.e., $\beta_1=2$. At this semi-Dirac point it can be shown that $D(E)$ loses its linear energy dependence and transforms to $D(E)\propto \sqrt{|E|}$, as highlighted in red in Fig.~\ref{fig:figsemi2}(c), in accordance to other works~\cite{Hasegawa2006,Wunsch2008,Adroguer2016}. 
    
    \item For $\beta_1>2$, the bands do not touch anymore, opening thus a band-gap around the Dirac energy, as clearly shown by $D(E)$.
\end{itemize}
 
  \begin{figure}[tb]
        \centering
        
        \includegraphics[width=0.45\textwidth]{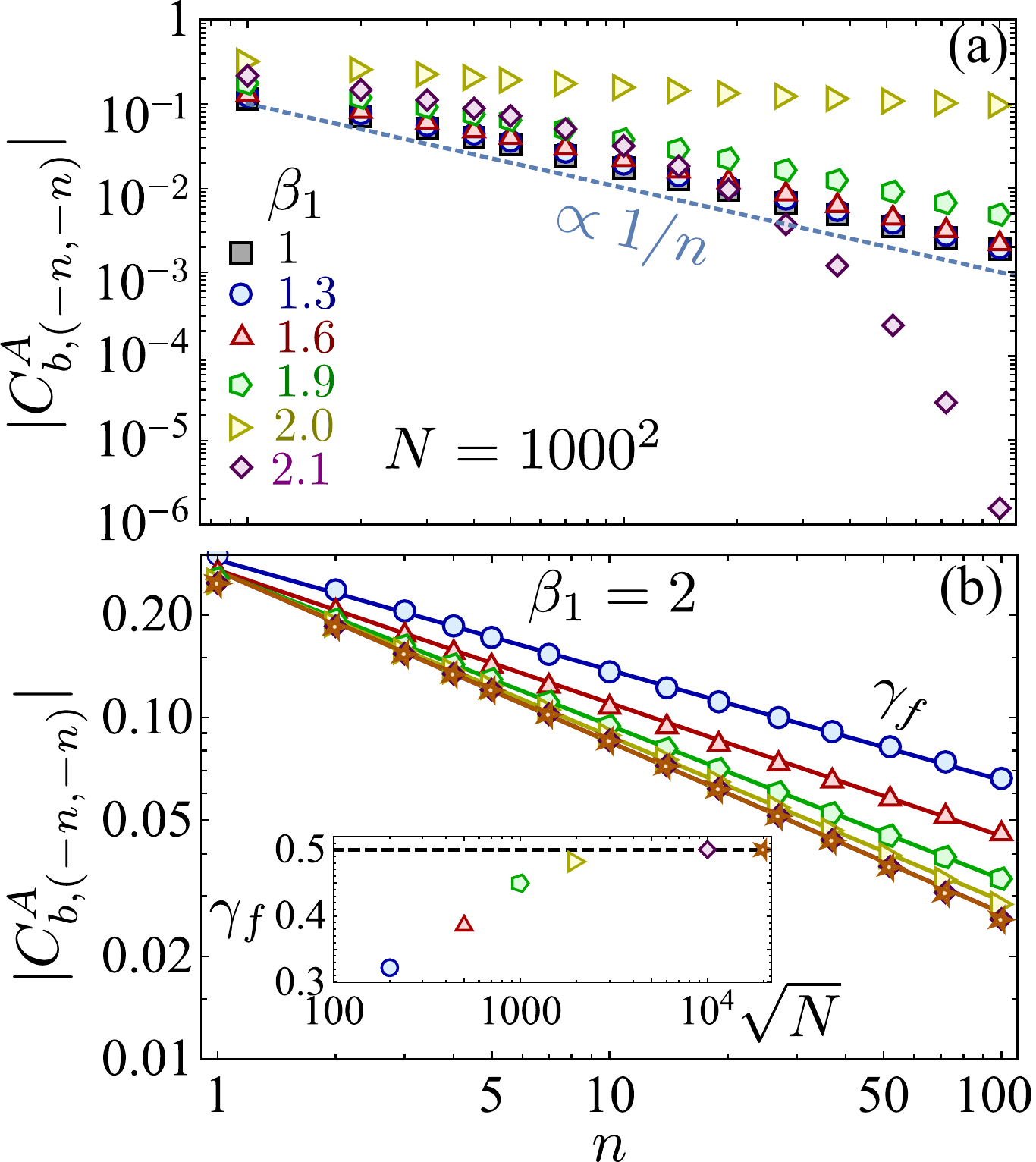}
        \caption{(a) Diagonal cut of the spatial shape of the quasi-bound state $|C^A_{b,-(n,n)}|$ for a bath size $N=1000^2$ and several $\beta_1$ as depicted in the legend. Dashed blue line is a guide to the eye of the scaling $1/n$. (b) Same diagonal cut for fixed $\beta_1=2$, and increasing system size $N$ as depicted in the legend. Solid lines are numerical fits to a power-law decay $A/n^{\gamma_f}$. In the inset, we plot the exponent $\gamma_f$ obtained from the numerical fit of the asymptotic shape of the quasi-bound state. Dashed black line is set at $0.5$, which is the value where $\gamma_f$ converges for large system sizes.}
        \label{fig:figsemi3}
    \end{figure}
    
    We expect that this transition from a gapless Dirac point to a gapped one passing through the semi-Dirac situation, also translates in important changes of the quasi-bound state appearing for the isotropic case. Since $\Sigma^\alpha_e(0)=0$ for all $\beta_1$, it is guaranteed that a bound state exists at $E_\mathrm{BS}=0$ (see Eq.~\eqref{eq:solBS}) for the whole anisotropy range. We can then characterize numerically its spatial shape by solving Eqs.~\eqref{eq:cAA}-\eqref{eq:cAB} for several $\beta_1$ and a fixed system size. The result of this analysis is shown in Fig.~\ref{fig:figsemi2}(a-f) where we plot the bound state shape $C^A_{b,\nn}$ for several $\beta_1$ ranging from $\beta_1=1$ (isotropic case, in panel (a)) to $\beta_1=2.1$ (gapped situation, in panel (b)), passing through the semi-Dirac situation $\beta_1=2$ (in panel (e)). From this analysis we can extract several conclusions:
    \begin{itemize}
        \item As it occurs in the case of 3D photonic Weyl points~\cite{Garcia-Elcano2020,Garcia-Elcano2021}, since the density of states maintains its singular nature for $\beta_1\in [1,2]$ one can modify the shape of the bound state without altering its power-law dependence. The main change of the spatial shape is due to the modification of the interference factor of $I_{\mathbf{K},\mathbf{K'}}$ due to the movement of the Dirac points $\mathbf{K}^{(')}$ with the anisotropy parameter $\beta_1$.  In fact, the asymptotic decay for large distances still shows a similar $1/r$ dependence:
        \begin{align}
        |I_{\mathbf{K}^{(')}}(\pp)|\propto \frac{1}{\sqrt{v_1^2\tilde{p}_1^2+v_2^2\tilde{p}_2^2}},\label{eq:Ikmod}
    \end{align} 
    with a modification compared to the isotropic case of Eq.~\eqref{eq:Ik} provided by the effective velocities $v_{i}$ in the different directions. This perseverance of the $1/r$ decay in this regime can be better observed in Fig.~\ref{fig:figsemi3}(a) where we plot a line cut of the quasi-bound-state $|C^A_{b,(-n,-n)}|$ for several $\beta_1$'s. There, we observe how for the $\beta_1$'s in that regime, i.e., $\beta_1=1,1.3,1.6,1.9$, the asymptotic scaling of the quasi-bound state follows perfectly a $1/n$-law, as plotted in dashed blue line. Something similar can be found along other directions (up to the oscillatory factors introduced by the interference of the two integrals of Eq.~\eqref{eq:Gijcont}).
    
    \item This situation changes dramatically at the semi-Dirac point $\beta_2=2$, see Fig.~\ref{fig:figsemi2}(e), where the interference between $I_{\mathbf{K},\mathbf{K'}}$ has more striking consequences. On the one hand, it leads to a directional bound state, that is, the quasi-bound state localizes mostly along one particular direction, that is, the $(-n,-n)$ direction where the $v_2$ velocity vanishes. Evidently, a different choice of the anisotropy direction in the hopping model of Fig.~\ref{fig:scheme}(b) will result in a different localization direction, but will not alter the rest of the conclusions. 
    
    Besides this directionality, the localization of the bound state can be shown to be longer ranged. This is more clear when we look at a line cut along this direction, as shown in Fig.~\ref{fig:figsemi3}(a), where we observe how the bound state decay with $\beta_1=2$ (yellow triangles) is significantly longer ranged than the other ones. In fact, one can make a numerical fit to a power-law decay $\propto 1/n^{\gamma_f}$ obtaining a value $\gamma_f\approx 0.45$ for the system size chosen in Fig.~\ref{fig:figsemi3}(a), i.e., $N=1000^2$. Through a numerical analysis we demonstrate that this exponent $\gamma_f$ depends on the system size. This is shown in Fig.~\ref{fig:figsemi3}(b) where we plot the quasi-bound state cut for $\beta_1=2$ and several system sizes (markers), together with its corresponding fittings (solid lines). The different slope in the logarithmic scale indicates a different exponent of the power-law decay.  To make it more evident, we complement this figure with the inset, where we plot the exponent of the power-law, $\gamma_f$, obtained from the numerical fit. There, we observe how $\gamma_f$ ranges from $0.3$ for the smallest system size considered, i.e., $N=200^2$, to $0.5$, which is the value it converges to in the limit of very large systems. We want to note that both the directional behaviour and $1/\sqrt{x}$ dependence resembles the results related to electronic and photonic transport found in the literature~\cite{Dutreix2013,Real2020}.
    
    \item Finally, when $\beta_1>2$ the quasi bound-state transforms into a standard one recovering its exponential decay dependence, as clearly shown in purple rombus in Fig.~\ref{fig:figsemi3}(a). Remarkably, though, this exponential localization preserves the one-dimensional directional behaviour from the semi-Dirac dispersion, as shown clearly in Fig.~\ref{fig:figsemi2}(f).
    \end{itemize}
    
After this exploration of the modification of the (quasi)-bound state spatial shape, a natural question arises: will the dynamical features be similar to the ones of the isotropic model explained in the previous section, or are there important qualitative differences? This is explored in Fig.~\ref{fig:figsemi4} by plotting the numerically calculated excited state population, $|C_e(t)|^2$ of a single initially excited emitter coupled to a bath with strength $g=0.1J_1$, energy tuned to the Dirac point, $\Delta=0$, and for increasing system sizes, $N$, depicted in the different colors. For comparison, we plot in the different panels (a-f) the dynamics for the same $\beta_1$ parameters chosen in Fig.~\ref{fig:figsemi2}. Besides, we perform the simulation using a bath with periodic boundary conditions to compare with the results of Ref.~\cite{Gonzalez-Tudela2018}. On top of the dynamics we plot in dashed lines the expected steady-state value, $|R_0|^2$, obtained from calculating the overlap with the quasi-bound state given by Eq.~\eqref{eq:R0} but replacing $\omega_\mathrm{iso}(\kk)\rightarrow \omega_1(\kk)$. From this analysis, we can extract several conclusions:

        \begin{figure}[tb!]
        \centering
        
        \includegraphics[width=0.45\textwidth]{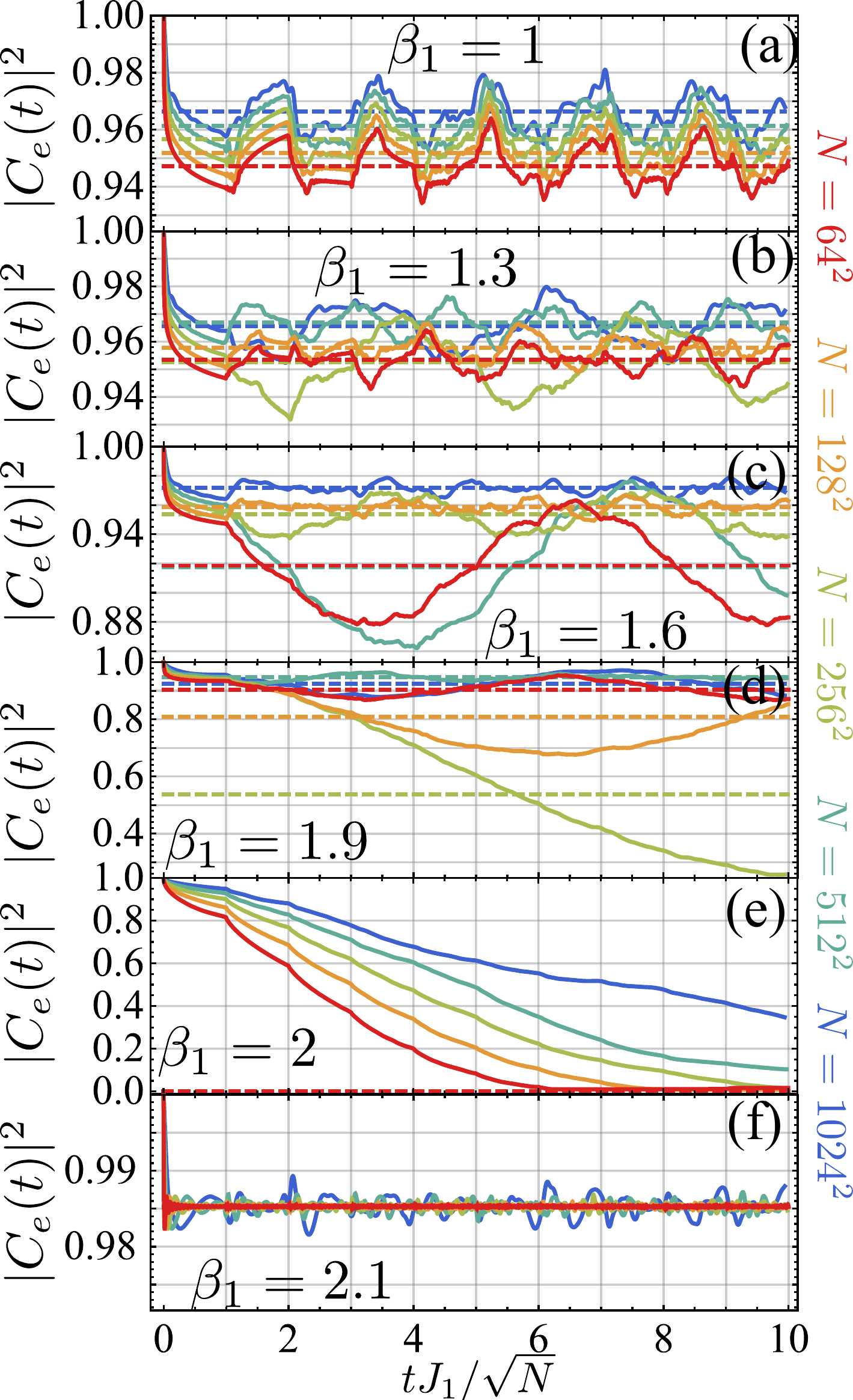}
        \caption{(a-f) Excited state population, $|C_e(t)|^2$, for an initially excited emitter for the same $\beta_1$ parameters of Fig.~\ref{fig:figsemi2}. In each panel, we plot the dynamics for increasing system sizes in different colors, as depicted in the legend. The bath is simulated with periodic boundary conditions and the emitter-bath coupling strength is $g=0.1J_1$. In dashed lines we plot the expected steady-state values, $|R_0|^2$, obtained from the overlap with the quasi-bound state given by Eq.~\eqref{eq:R0}.}
        \label{fig:figsemi4}
    \end{figure}
    
\begin{itemize}
    \item For $\beta_1\in [1,2]$, and for times in which the photonic excitations do not \emph{see} the border, i.e., $t J_1/N<1$, all $\beta_1$ show a similar non-exponential decay dynamics which resembles the one of the isotropic case. Thus, for an infinite system, $N\rightarrow \infty$, we expect then the dynamics to be dominated purely by the non-Markovian dynamics, $C_\mathrm{NM}(t)$ in Eq.~\eqref{eq:cetiso}, induced by the non-analytical behaviour of the density of states around the Dirac energy.
    
    \item For $\beta_1\in(1,2)$ and \emph{long-enough} times, i.e., $t J_1/N>1$, the decay quenches and the population starts to oscillate around a steady-state value, like in the isotropic case. This value is determined by the overlap of the quasi-bound state with the emitter, $|R_0|^2$. As expected, $R_0$ depends now not only on system size, $N$, but also on the anisotropy parameter $\beta_1$, i.e., $R_0\equiv R_0(N,\beta_1)$. However, as we show in Figs.~\ref{fig:figsemi4}(b-e), an important difference arises with respect to the isotropic case: for fixed $\beta_1$, $R_0$ does not monotonically decrease to $0$ with $N$, but rather with some oscillations that depend on the $\beta_1$ parameter. The same occurs with the dependence with $\beta_1$ for fixed system size. As a general rule, we find that $R_0$ tends to zero as $\beta_1$ gets closer to the critical value, $\beta_1\rightarrow 2$, as indicated by the larger oscillations found in Figs.~\ref{fig:figsemi4}(c-d).
    
    \item At the critical point, $\beta_1=2$, we find that $R_0(\beta_1=2,N)\equiv 0$ for all evaluated system sizes. The underlying reason is that for this case a zero-energy extended mode appears, for the boundary conditions chosen, that resonantly couples the emitter. Thus, the emitter excitation gets resonantly transferred to such extended mode, as shown in Fig.~\ref{fig:figsemi4}(e), and it ultimately comes back coherently to the emitter at sufficiently long times (not shown).
    
    \item Finally, for $\beta_1>2$ one recovers the dynamical features of the standard qubit-photon bound states appearing in conventional band-gaps~\cite{john94a}: that is, the excitation decays initially up to a constant value $|R_0|^2\approx 1$, and independent of system size. This is very clear in Fig.~\ref{fig:figsemi4}(e) where we see the collapse of all the lines for the dynamics of different system sizes.
\end{itemize}
    
Summing up, we have shown that the anisotropy of Dirac lattices provides a very useful knob to tune the spatial shape of the (quasi)-bound state, and thus, on the effective coherent interactions, $G_{ij}^{\alpha\beta}$ appearing in Eq.~\eqref{eq:effH}. For the regime $\beta_1\in(1,2)$ one can tune the spatial shape without losing its power-law behaviour, something only found so far for three-dimensional topological baths~\cite{Garcia-Elcano2020,Garcia-Elcano2021}. For $\beta_1>2$ one can obtain a directional and longer-ranged interaction ($1/\sqrt{r}$), corrected by an exponential decay length which decreases with the value of $\beta_1$. At the critical point $\beta_1=2$, however, the overlap with the quasi-bound state goes to zero, such that it can not eventually mediate the interactions $G_{ij}^{\alpha\beta}$ for very long times.

    \begin{figure*}[tb]
        \centering
        \includegraphics[width=\textwidth]{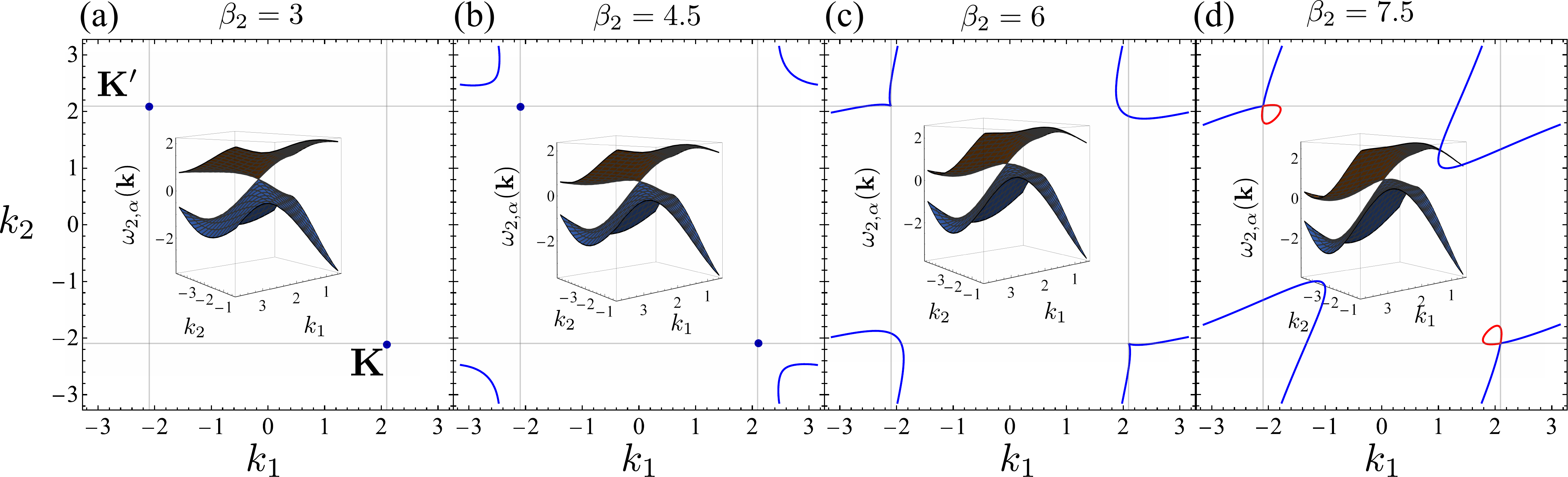}
        \caption{(a-d) Evolution of the nodal lines, $\omega_{2,\alpha}(\kk)=\tilde{E}_D$ for a tilted Dirac photonic bath for several $\beta_2$ as indicated in the legend. The rest of the parameters of the model are $J_2=0.1J_1, \beta_1=1$. Here,  $\tilde{E}_D$ is the shifted Dirac energy of the model defined by Eq.~\eqref{eq:eDirac}. We plot in blue/red the nodal lines of the upper/lower band. At the middle, we make a three-dimensional plot of the energy dispersion, $\omega_{2,\alpha}(\kk)$, around the Dirac point $\mathbf{K}$ of the model.}
        \label{fig:tilt1}
    \end{figure*}
\section{Photon-mediated interactions in tilted Dirac points \label{sec:tilted}}

For completeness in the study of the possible Dirac-like dispersions that can appear, in this section we will consider a generalization of the hopping model studied in the previous section by including anisotropic next-nearest neighbour hoppings, i.e., making $J_2\neq 0$ and $\beta_2\neq 1$ (see Fig.~\ref{fig:scheme}(b)). The reason is that this simplified model will allow us to tilt the energy dispersion~\cite{Goerbig2008,Lu2016,Katayama2006,Hirata2016}, and explore the quantum optical phenomena induced by the so-called Type II and III Dirac points~\cite{Soluyanov2015,Xu2015a,Deng2016,Huang2016,Noh2017,Huang2018,Liu2017b,Perrot2018,Milicevic2019a,Mizoguchi2020,Jin2020, Kim2020a}, which occur when a critical tilting leads to a linear band-touching along the Brillouin zone, instead of a single point as in the standard (Type I) Dirac points.

In the $2\times 2$ version of the bath Hamiltonian, $\tilde{H}_B(\kk)$, the addition of next-nearest neighbour hoppings enters in the diagonal part $h_I(\kk)\neq 0$ written in Eq.~\eqref{eq:hIk}. Thus, the energy dispersion will be given by two bands with energies:
\begin{align}
    \omega_{2,u/l}(\kk)=h_I(\kk)\pm \omega_1(\kk)\,.\label{eq:endisp2}
\end{align}

Note that $h_I(\kk)$ provides a $\kk$-dependent shift common to the upper and lower bands. Thus, the appearance (or not) of band-touching points is still dominated by the $\omega_1(\kk)$ studied in the previous section. The main effect of $h_I(\kk)$ is to introduce both a tilting and a shift of the energy dispersion around the Dirac points. This is more clear when expanding the Hamiltonian around the Dirac points:
\begin{align}
        \Tilde{H}_B(\mathbf{K}/\mathbf{K}^{'}+\mathbf{q})\approx \tilde{E}_D+(v_{0,1}q_1\mp v_{0,2}q_2)\mathbf{1}+ \\
        +v_{1}\sigma_y q_1\mp v_{2}\sigma_x q_2, \label{eq:expHbaniso}
\end{align}
for $|\qq_i|\ll 1$, where $\tilde{E}_D$ is the shifted energy of the Dirac points given by:
\begin{align}
    \tilde{E}_D=J_2\left[2\beta_2+\beta_1(2-\beta_1\beta_2)\right]\,,\label{eq:eDirac}
\end{align}
$v_{1/2}$ the slopes around the Dirac energy defined in Eq.~\eqref{eq:v1v2}, and $v_{0,1/2}$ the new effective diagonal velocities, which can be shown to be given by:
\begin{align}
v_{0,1}=0\,,v_{0,2}=2\frac{J_2}{J_1} v_2\left(\beta_1\beta_2-1\right)\,.\label{eq:v01v02}
\end{align}

As explained in Refs.~\cite{Goerbig2008,Milicevic2019a}, with these velocities, $v_{0,i},v_i$, one can define a parameter, $\tilde{v}$:
\begin{align}
        \tilde{v}&=\sqrt{\left(\frac{v_{01}}{v_1}\right)^2+\left(\frac{v_{02}}{v_2}\right)^2}=\frac{2J_2}{J_1}\,\lvert \beta_1\beta_2-1\rvert\,,\label{eq:tilting}
    \end{align}
that characterizes the degree of tilting of the lattice, and allows to classify the different type of Dirac band touching that occur (see, e.g., Ref.~\cite{Milicevic2019a} for an explanation of the classification). For $\tilde{v}\in(0,1)$ the energy dispersion is tilted but still has a single-point band-touchings (Type I Dirac points). At the critical value $\tilde{v}=1$, the energy dispersion undergoes a transition from a single-point band touching to a combination of linear band touching and flat band (Type III). Finally, when $\tilde{v}>1$ the band is over-tilted showing the so-called Type II Dirac points characterized by nodal line crossings.

In Figs.~\ref{fig:tilt1}(a-d) we show an example of this transition for the energy dispersions $\omega_{2,u/l}(\kk)$ of Eq.~\eqref{eq:endisp2}. For this plot, we fix $J_2=0.1 J_1$, $\beta_1=1$, and move the anisotropy parameter for positive $\beta_2$ ranging from $\beta_2=3$ to $7.5$ in $1.5$ steps. From Eq.~\eqref{eq:tilting} and with this choice of parameters, it can be shown that the critical tilting $\tilde{v}=1$ should occur for $\beta_2=6$ (corresponding to Fig.~\ref{fig:tilt1}(c)). However, what happens in this anisotropic model is more intricate. For small enough anisotropy $\beta_2$  the energy dispersions are indeed tilted and only touch at a single point corresponding to the Dirac energy $\tilde{E}_D$, see Fig.~\ref{fig:tilt1}(a). For bigger $\beta_2$, though still smaller than the critical value for which $\tilde{v}=1$, the energy dispersion still have linear band touchings at the $\mathbf{K}^{(')}$ points, however, now there are additional $\kk$-modes from the upper band that become resonant with the Dirac energy, as shown in the corners of Fig.~\ref{fig:tilt1}(b). The latter will be important when we couple emitters since a point dipole generally couples to all $\kk$-modes which are resonant to the emitter's optical transition. At the critical value, $\beta_2=6$ for this choice of parameters, the energy dispersion indeed undergoes a transition to a linear band-touching around $\mathbf{K}^{(')}$ points, although again accompanied by other resonant modes at the same energy (see Fig.~\ref{fig:tilt1}(c)). Finally, for larger $\beta_2$, the energy bands are overtilted and there are resonant modes at the Dirac energy for both the upper/lower bands, as it is expected for type II Dirac points.

       \begin{figure}[tb]
        \centering
        \includegraphics[width=0.44\textwidth]{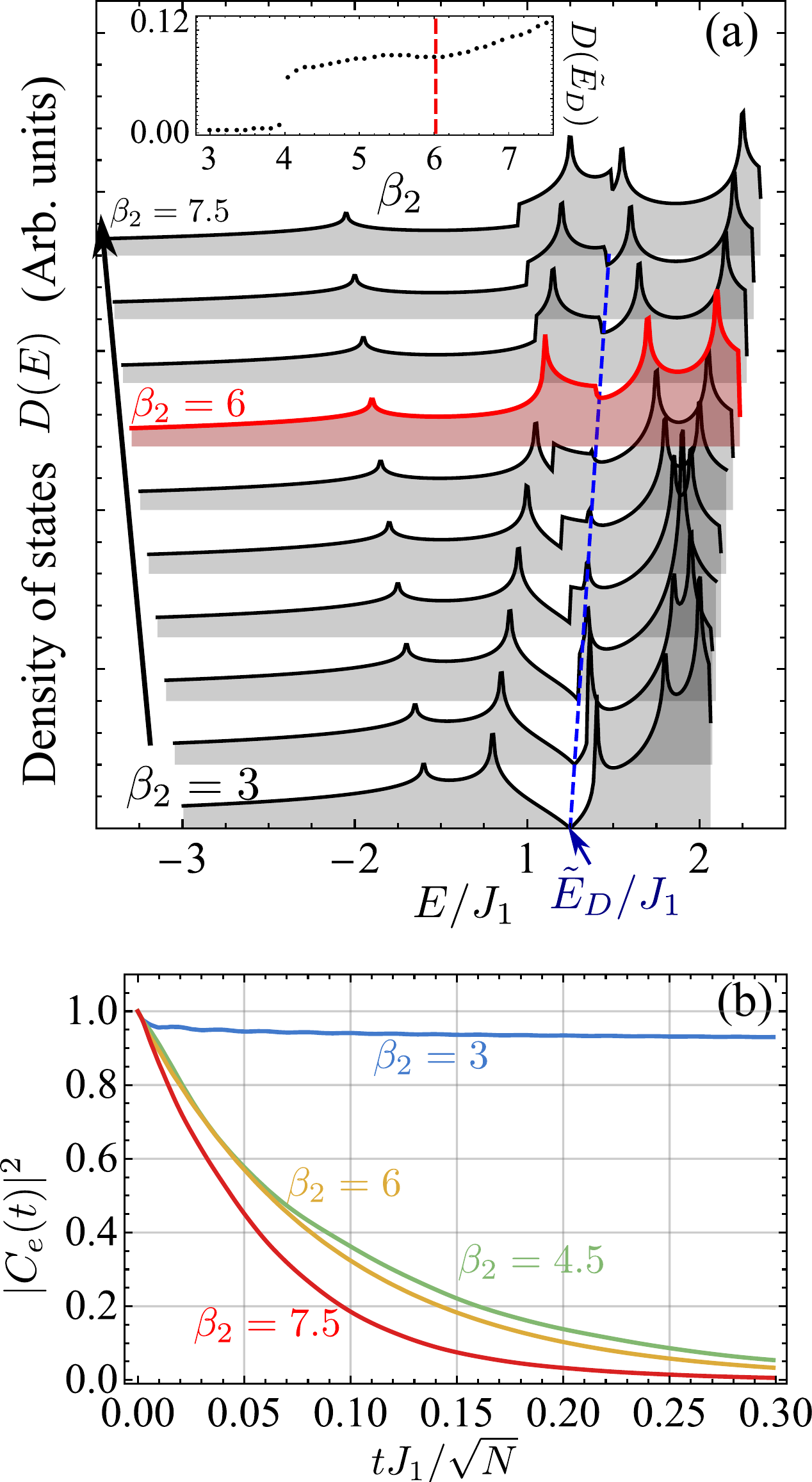}
        \caption{(a) Density of states for the tilted Dirac photonic bath with $J_2=0.1J_1, \beta_1=1$ for several $\beta_2$ ranging from $3$ to $7.5$ in steps of $\Delta\beta_2=0.5$. We highlight in red the case $\beta_2=6$ which is the critical tilting parameters $\tilde{v}$ predicted by Eq.~\eqref{eq:tilting}. Dashed blue line indicates the position of the Dirac energy $\tilde{E}_D$ due to the shift explained in Eq.~\eqref{eq:eDirac}. Inset: $D(\tilde{E}_D)$ as a function of $\beta_2$ parameter.(b) Excited state population dynamics $|C_e(t)|^2$ for an emitter coupled with strength $g=0.1J_1$ to the central $A$ site of a photonic bath with $J_2=0.1J_1, \beta_1=1$, $N=1024^2$, and the $\beta_2$'s plotted in Fig.~\ref{fig:tilt1}. The emitter is always tuned to the Dirac energy position $\tilde{E}_D$ so that it always probes the Dirac energy spectrum}
        \label{fig:tilt2}
    \end{figure}
    
Before considering what happens when an emitter is coupled to this bath, it is instructive to understand first how this transition is manifested in the density of states of the bath, $D(E)$, that we know plays a very relevant role in the emitter's dynamics. In Fig.~\ref{fig:tilt2}(a) we study the evolution of $D(E)$ as a function of $\beta_2$ for the same parameters than Fig.~\ref{fig:tilt1}. We take a range of $\beta_2\in (3,7.5)$ in steps of $0.5$, which includes the transition through the critical value $\beta_2=6$, highlighted in red in Fig.~\ref{fig:tilt2}(a). There, we observe several features: i) As expected from Eq.~\eqref{eq:eDirac}, the Dirac point shifts to larger values as $\beta_2$ increases. We highlight this movement in dashed blue lines. There, we observe that for $\beta_2\lessapprox 4$, the density of states at the Dirac energy is approximately zero, as expected from type I Dirac points. However, for $\beta_2\gtrapprox 4$ the density of states starts to acquire a non-zero value. This is more clear at the inset panel of Fig.~\ref{fig:tilt2}(a), where we plot the value of the density of states precisely at the Dirac energy, $D(\tilde{E}_D)$. The reason for this finite value at the density of states are precisely the $\kk$-resonant modes that start to appear at other areas of the Brillouin zone far from the linear band-touching at the $\mathbf{K}^{(')}$-points. As $\beta_2$ goes to a critical point, one of the Van-Hove singularities of the density of states approaches and crosses the Dirac point, which is one of the signatures of Type III-Dirac points. For $\beta_2>6$, the density of states around the Dirac point is also finite, as expected for Type II Dirac points, and increasing since more $\kk$-modes are available for the Dirac energy $\mathrm{E}_D$. 

Let us now see what happens when emitters couple to these type of photonic baths, and how the transition through the critical Type III Dirac point affects quantum optical phenomena. As for semi-Dirac points, we will use spontaneous emission of a single initially excited emitter as it will allow to unravel phenomena which has to do both with the localization of photon emission, as it is expected in Type I Dirac points, and with the emission of propagating photons, as it is expected for Type II-III Dirac points. We start by plotting in Fig.~\ref{fig:tilt2}(b) the excited state population of an emitter coupled, with strength $g=0.1J_1$ and energy $\Delta=\tilde{E}_D$, to the central $A$-site of a photonic lattice with the parameters of Fig.~\ref{fig:tilt1}(a-d). There, we observe how the slow non-Markovian relaxation expected for Type I Dirac points is only observed for the smallest $\beta_2$ chosen, where there are no additional $\kk$-resonant modes. For the other cases, the emitter dynamics displays the standard exponential relaxation to the ground state, with a decay rate proportional to the value of the density of states at the emitter's energy, $D(\tilde{E}_D)$. Thus, the presence of other resonant $\kk$-modes does not allow to observe any strong signature of the crossing of the critical point in the quantum emitter dynamics.

       \begin{figure}[tb]
        \centering
        \includegraphics[width=0.44\textwidth]{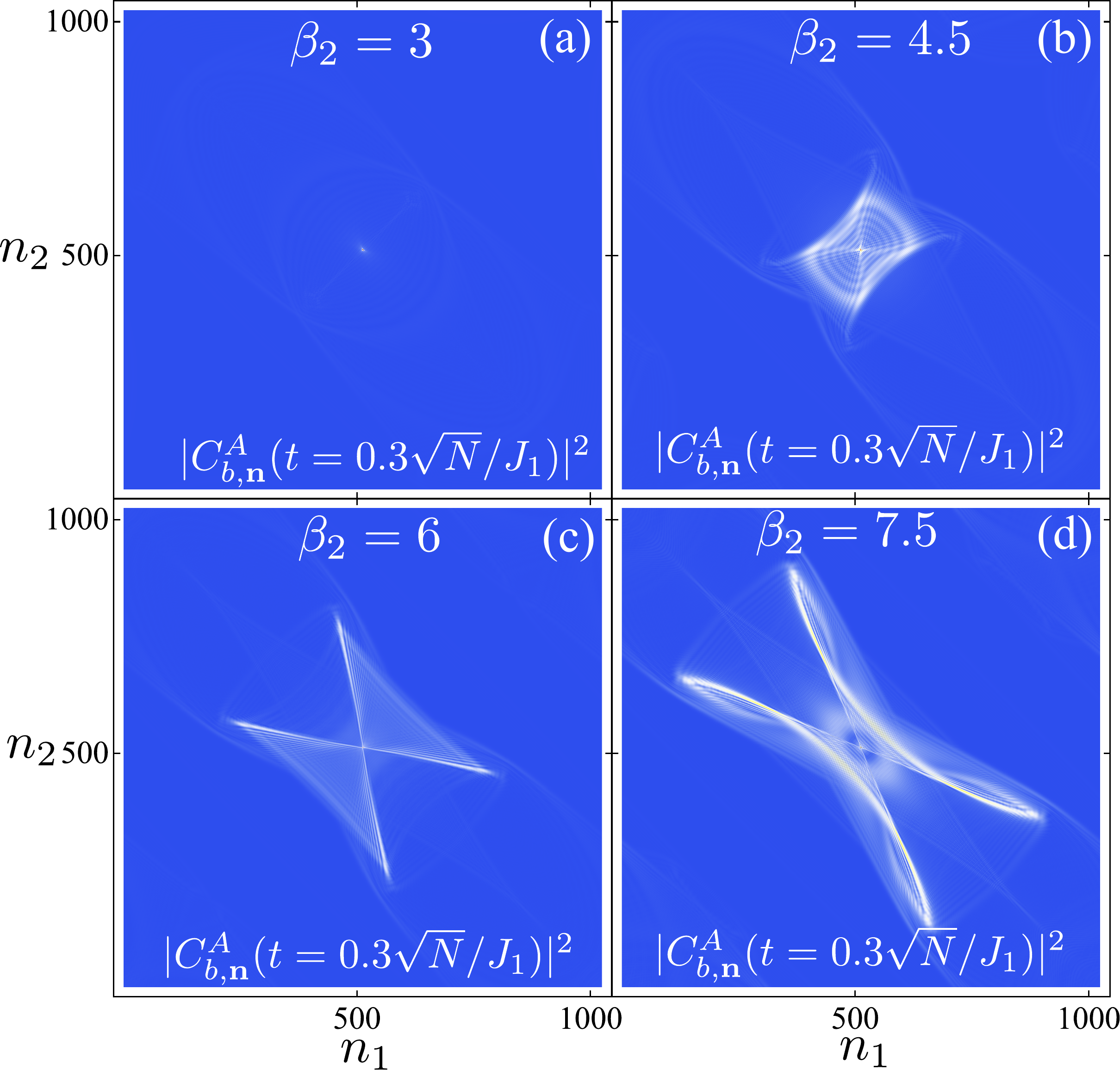}
        \caption{(a-d) Photonic bath population in the $b$ sites (the $a$ sites show a similar behaviour) after a time $t=0.3\sqrt{N}/J_1$ of a quantum emitter coupled to the central $A$-site of a photonic bath with the parameters of Fig.~\ref{fig:tilt2}(b). Different panels correspond to different values of $\beta_2$ as depicted in the legend.}
        \label{fig:tilt3}
    \end{figure}
    
On the contrary, one can indeed observe some signatures by looking the radiation pattern in the bath, given by the wavefunctions $C^A_{a/b,\mathrm{n}}(t)$ of Eq.~\eqref{eq:ansatz}. This is what we plot in Fig.~\ref{fig:tilt3} by choosing the same parameters of Fig.~\ref{fig:tilt2}(b) and looking at the wavefunction at the time $t=0.3\sqrt{N}/J_1$. We only plot the population in the $b$-modes since the one of the $a$ sites displays a qualitatively similar behaviour (except for $\beta_2=3$ where the $a$ modes shows no population). In these panels, we can observe how for $\beta_2\lessapprox 4$ (Fig.~\ref{fig:tilt3}a) the light becomes strongly localized around the emitter, being thus dominated by the quasi-bound state physics discussed in section~\ref{sec:anisotropy}. When the other $\kk$-modes become resonant, like in $\beta_2=4.5$~(Fig.~\ref{fig:tilt3}b), the population into the bath is a superposition between the quasi-bound state localization and the propagating emission coming from the other modes available. At the critical point~(Fig.~\ref{fig:tilt3}c), the emission becomes strongly directional due to the presence of a Van-Hove singularity around the Dirac point~\cite{Gonzalez-Tudela2017b,Gonzalez-Tudela2017a,galve17a}. As we will see next, this directional emission is a strong indication of the crossing of the critical point of Type III Dirac points. Here, we want to note that one also gets emission from the other $\kk$-resonant modes, however, since the density of states associated to the Van-Hove singularity is typically much larger, it eventually dominates the emission. Finally, for $\beta_2=7.5$ corresponding to the Type-II situation (see Fig.~\ref{fig:tilt3}d), the emission starts to lose its directional character due to the increasing number of $\kk$-modes that become resonant with the emitter's energy (see Fig.~\ref{fig:tilt1}d).

    \begin{figure*}
        \centering
        \includegraphics[width=0.94\textwidth]{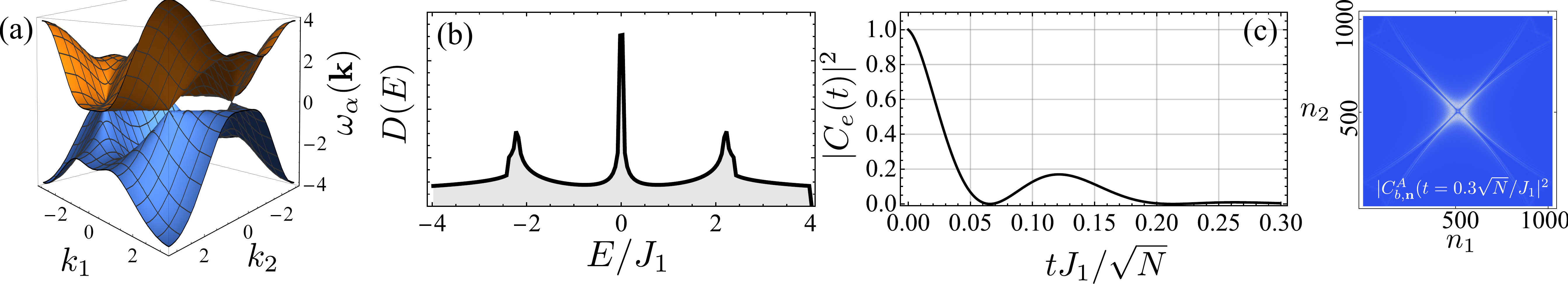}
        \caption{(a) Energy dispersion $\omega_{u/l}(\kk)$ of the model described in Eq.~\eqref{eq:HBmizo}, built in Ref.~\cite{Mizoguchi2020} to show a perfect Type III energy dispersion. Parameters: $J_2=0.3J_1$. (b) Associated density of states, $D(E)$, for the model and parameters of panel (a). The Type-III energy dispersion leads a large Van-Hove singularity at the Dirac energy ($0$). (c) Excited state population of an initially excited emitter coupled to the middle A site of a photonic bath described by the Hamiltonian of Eq.~\eqref{eq:HBmizo}. Parameters: $J_2=0.3J_1$, $N=1024^2$, $g=0.1$, $\Delta=0$. (d) Bath population (in the $b$ modes) after the de-excitation of the emitter described in panel (c). The population of the $a$ sites show a qualitatively similar behaviour.}
        \label{fig:figmizoguchi}
    \end{figure*}
   
As we have just seen, for the simple model we have chosen the type I-III-II transition is hindered by the presence of additional $\kk$-modes at the emitter's energy. A natural question then arises: how will spontaneous emission look like if the emitter could couple perfectly only to $\kk$ modes where the linear band-touching appear? One option to obtain this would be to keep the same photonic lattice model, but get of rid of the coupling to these spurious modes. This could be done, for example, using giant atoms~\cite{kockum18a,FriskKockum2021,Kannan2020,Wang2021a,Gonzalez-Tudela2019b} and designing their couplings such that the emitter only couple to the linear band-touching modes. Here, however, we will follow a cleaner approach just for the sake of illustration. We will take an alternative photonic lattice model proposed in Ref.~\cite{Mizoguchi2020}, especially designed to have a perfect Type-III energy dispersion. The model can be written as a bipartite bath as follows:
\begin{align}
        \tilde{H}_B(\kk)=\begin{pmatrix}
        d(\kk)+a(\kk) & a(\kk) \\
        a(\kk) & a(\kk)-d(\kk)
        \end{pmatrix},\label{eq:HBmizo}
\end{align}
where $d(\kk)=2J_1 (\cos k_1+\cos k_2)$ and $a(\kk)=2J_2 (\cos k_1-\cos k_2)$. This model has two bands $\omega_{u/l}(\kk)=a(\kk)\pm \sqrt{a(\kk)^2+d(\kk)^2}$. Since $d(\kk)$ vanishes at $k_1\pm (\mp) k_2=\pm\pi$, and $a(\kk)$ vanishes when $k_1=\pm k_2$, the type III Dirac cones appear at the intersection of those lines, i.e, at the points $\kk=(\pm \pi/2, \pm \pi/2)$, as can be seen in Fig.~\ref{fig:figmizoguchi}(a) where we plotted $\omega_{u/l}(\kk)$ for $J_2=0.3J_1$. 

Apart from having a perfect singular band-touching, the energy dispersion along the straight lines $k_1\pm (\mp) k_2=\pm\pi$ are perfectly flat, which ultimately results in a highly divergent density of states, i.e., Van-Hove singularity, at the Dirac energy. This is clearly seen in Fig.~\ref{fig:figmizoguchi}(b), where we plot $D(E)$ for this model with $J_2=0.3J_1$. This divergent density of states has strong impact in the quantum emitter dynamics, that we plot in Fig.~\ref{fig:figmizoguchi}(c), for an emitter coupled to the central $A$ site of a photonic lattice with $J_2=0.3J_1$ and $N=1024^2$ sites. Differently from the critical point dynamics of Fig.~\ref{fig:tilt2}(b), the emitter now displays a strong non-Markovian relaxation showing even some reversible oscillations. Besides, these oscillations are also accompanied by a highly directional emission into the bath, as plotted in Fig.~\ref{fig:figmizoguchi}(d). Thus, both the reversible dynamics and directional emission can be considered as smoking guns of the critical Dirac transition.

\section{Anisotropic and tilted energy dispersions in subwavelength atomic arrays \label{sec:implementation}}

\begin{figure*}[!tb]
  \centering
  \includegraphics[width=\textwidth]{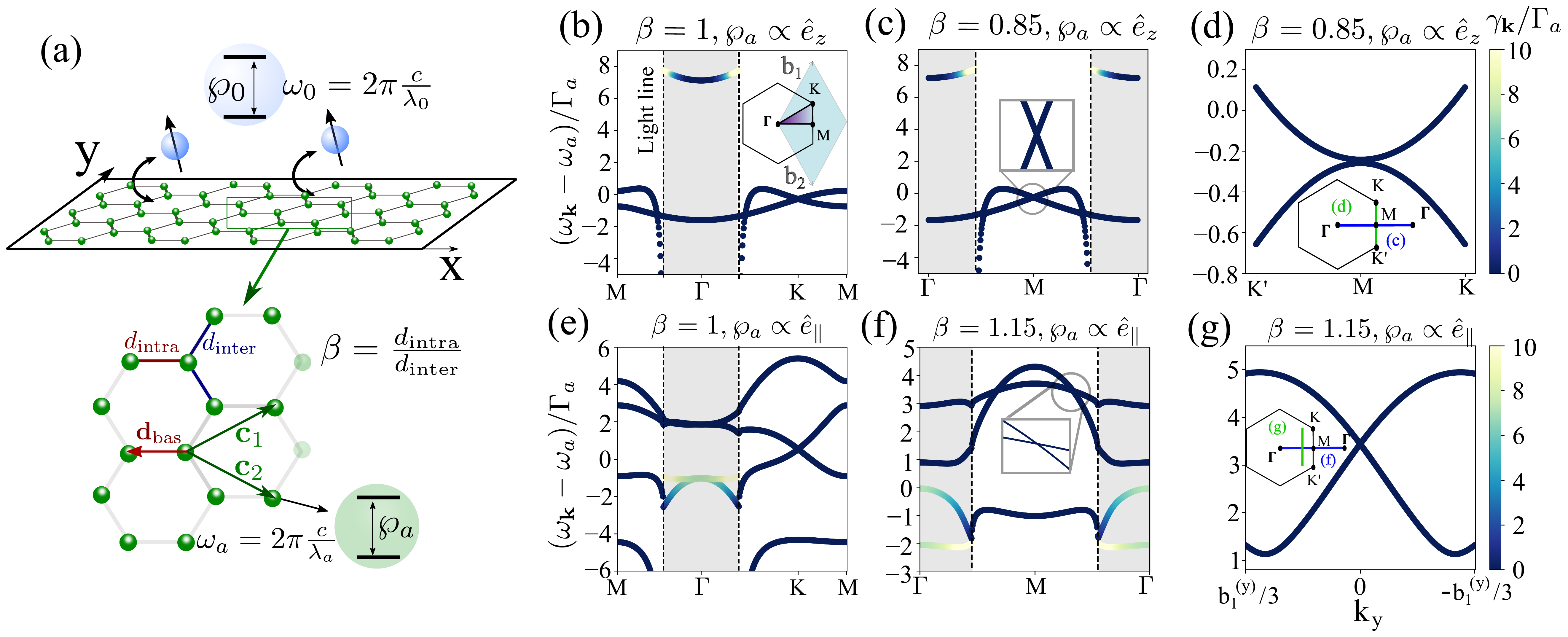}
  \caption{(a) Subwavelength atomic array with tunable honeycomb geometry hosts subradiant atomic excitations with non-trivial Dirac energy dispersions $\omega(\kk)$. When additional atoms are placed nearby the structure, they can exchange interactions with these subradiant modes, mimicking light-matter couplings. We tune the honeycomb geometry keeping the primitive lattice vectors fixed: $\cc_{1/2}=d\sqrt{3}/2(\sqrt{3}\hat{\bf{e}}_x\pm \hat{\bf{e}}_y)$, and modifying the ratio between the intra/inter-cell distance given by $\beta=d_\mathrm{intra}/d_\mathrm{inter}$. Like this, one generates lattices with uniaxial anisotropy and tune the energy dispersions of the subradiant excitations. (b) Dispersion relation of an isotropic honeycomb lattice, $\beta=1$, with $d_\mathrm{inter}=d_\mathrm{intra}=0.15\lambda_a$, for out of plane modes. (c,d) We keep $d=0.15\lambda_a$ fixed but change $\beta<1$, and plot the dispersion of the out of plane modes for this anisotropic lattice. A semi-Dirac point emerges at $M$ for $\beta=0.85$, as shown by plotting the two paths in the Brillouin zone sketched in the inset: while the dispersion is linear along $k_x$, it is quadratic along $k_y$. (e) Dispersion relation of the in plane modes of the same isotropic honeycomb lattice as in (b). (f,g) We now tune $\beta=1.15>1$, and show how a tilted Dirac cone emerges for the in-plane modes of this anisotropic lattice. We plot the dispersion along $k_x$ (f) and $k_y$ (g) following the  two different paths in the Brillouin zone sketched in the inset. In all panels the color scale represents the decay rate, $\gamma_\kk$, of the modes as indicated in the legend. }
  \label{fig:subwavelenth}
\end{figure*}

In this section, we discuss a possible implementation to observe the phenomena predicted along this manuscript based on the recent proposals for exploring quantum optical phenomena using subwavelength atomic arrays~\cite{Masson2020,Patti2021,Brechtelsbauer2020}. The idea of these proposals is sketched in Fig.~\ref{fig:subwavelenth}(a): a few impurity atoms (in blue) are placed near a subwavelength atomic array (in green), so that they can couple to the subradiant atomic excitations that propagate within the array with an energy dispersion, $\omega(\kk)$, which depends on the geometry/atomic parameters~\cite{asenjogarcia17a,Asenjo-Garcia2019a,Zhang2019b,Bettles2016b,Wang2018,Bettles2016c,Bettles2020a,shahmoon17a,perczel17a,Perczel2017,bettles17a,Wild2018a}. Such subwavelength atomic arrays have been recently implemented using Rubidium atoms in square geometries~\cite{Rui2020} for inter-atomic distances $d/\lambda_a\approx 0.68$, with $\lambda_a$ being the wavelength associated to the optical transition. However, one can reach smaller distance regimes using Alkaline-Earth atoms~\cite{Masson2020,Patti2021,Brechtelsbauer2020}, since they feature a combination of short- and long-wavelength transitions~\cite{covey19a}. Besides, let us also note that other geometries, such as hexagonal or triangular lattices~\cite{struck11a,struck13a,soltan11a,Wintersperger2020}, have also been obtained using different laser configurations to generate the optical trapping potentials for the atomic arrays.

The goal of this section is to find a subwavelength atomic array configuration where the non-trivial energy dispersions discussed along this manuscript (semi- and tilted- Dirac points) appear for the subradiant modes of the array. For that purpose, we consider an hexagonal geometry as the one depicted in Fig.~\ref{fig:subwavelenth}(a): it is described by a two-site Bravais lattice with primitive vectors $\cc_{1/2}=d\sqrt{3}/2(\sqrt{3}\hat{\bf{e}}_x\pm \hat{\bf{e}}_y)$, and basis vector $\mathbf{d}_\mathrm{bas}=-d_\mathrm{intra}\hat{\mathbf{e}}_x$. For standard honeycomb lattices like the one depicted in the figure, the distance between the nearest-neighbours in all directions are equal, i.e.,  $d_\mathrm{intra}\equiv d$. However, in order to realise honeycomb lattices with uniaxial anisotropy we let $d_\mathrm{intra}\neq d$, while keeping the distance between the atoms within the same sublattice ($\sqrt{3}d$) fixed. We characterize the degree of anisotropy by a parameter defined as $\beta=d_\mathrm{intra}/d_\mathrm{inter}$, where $d_\mathrm{inter}$ is the intercell nearest neighbour distance as indicated in Fig.~\ref{fig:subwavelenth}(a). Hence, $\beta>(<)1$ represents an anisotropic lattice where the two atoms in the unit cell are pushed together (apart), respectively. In what follows, we will show how these anisotropic honeycomb atomic arrays display dispersion relations with semi-Dirac and tilted Dirac cones. 

We initially consider a 2D atomic array arranged in a honeycomb lattice $\beta=1$. Each atom is assumed to be a two level system with resonance frequency $\omega_a=2\pi c/\lambda_a$ and polarization dipole $\boldsymbol{\wp_a}$. The dynamics of this atomic array can be described within an stochastic wavefunction approach through the following effective non-Hermitian Hamiltonian~\cite{asenjogarcia17a,Perczel2017}:
\begin{align}
    \frac{H_{m}}{\hbar} = \sum_{j=1}^{N_A}\left(\omega_a-i\frac{\Gamma_a}{2}\right) \sigma_{ee}^j+  \sum_{i\neq j=1}^{N_A}\left(J_{ij}-i\frac{\Gamma_{ij}}{2}\right)\sigma_{eg}^i\sigma_{ge}^j\,,
\label{eq:H_eff}
\end{align}
where $j$ is a index running over all atoms in the metasurface ($N_A$), placed at positions $\rr_j$, and $\Gamma_a=|\boldsymbol{\wp_a}|^2\omega_a^3/(3\pi\hbar c^3)$ is the individual free-space decay rate. The coherent ($J_{ij}$) and incoherent ($\Gamma_{ij}$) photon-mediated interactions among emitters are given by the free space Green dyadic $\mathbf{G}_0(\rr_i-\rr_j)$~\cite{lehmberg70a,lehmberg70b}:
\begin{equation}
    J_{ij}-i\frac{\Gamma_{ij}}{2}=-\frac{3\pi \Gamma_a c}{\omega_a}\hat{\boldsymbol{\wp}}^*_{i} \cdot\mathbf{G}_0(\rr_i-\rr_j)\cdot\hat{\boldsymbol{\wp}}_{j}
\label{eq:Green_tensora}
\end{equation}
where $\hat{\boldsymbol{\wp}}_i=\boldsymbol{\wp}_{i}/|\boldsymbol{\wp}_i|$, and:
\begin{equation}
\begin{split}
    \mathbf{G}_0(\rr)= \frac{1}{4\pi}\left[\mathbb{1}+\frac{\nabla\otimes\nabla}{k_0^2}\right]\frac{e^{ik_0 |\rr|}}{|\rr|}\,.
\end{split}
\label{eq:Green_tensor1}
\end{equation}

Note that the couplings within the atomic array have two important differences with respect to the simplified coupled resonator models we consider along the manuscript: i) the photon-exchange interactions within the atomic array are long-ranged; ii) they depend on the polarization of the optical transitions. Thus, finding the same energy energy dispersions than in the nearest neighbour model is not guaranteed. In order to find such energy dispersions we restrict to the single-excitation subspace and the infinite size limits, where the eigenstates of the Hamiltonian $H_{m}$ are Bloch functions, 
\begin{equation}
 S^\dagger_\kk=\frac{1}{\sqrt{N}}\sum_{n=1}^N\sum_{m=1}^2 \sigma^{n,m}_{eg} e^{i\kk\cdot\rr_n}, 
\end{equation}  
where $\kk=(k_x,k_y)$ is the Bloch wavevector in the first Brillouin zone, and we explicitly take into account that we have a non-Bravais lattice, with the sums now running over $n$, up to the total number of unit cells ($N$) and over the two sites per unit cell ($m$). Obtaining the eigen-energies of the Bloch modes from the above Hamiltonian then reduces to diagonalising the following matrix~\cite{Perczel2017}:
\begin{align}
    \mathbf{M}^{\alpha\beta,\mu\nu}_\kk&=(\omega_a-i\frac{\Gamma_a}{2})\delta_{\alpha\beta}\delta_{\mu\nu} \\ &
    - \frac{3\pi \Gamma_a c}{\omega_a} \left[ \sum_{m=1}^2 \sum_{\bf{R}_n\neq0} e^{-i\kk\cdot\bf{R}_n} G^{\alpha\beta}_0(\bf{R}_n)\delta_{m\mu}\delta_{m\nu} \right. \nonumber \\ &
    + \sum_{\bf{R}_n} e^{-i\kk\cdot\bf{R}_n} G^{\alpha\beta}_0(\bf{R}_n+\mathbf{d}_\mathrm{bas})\delta_{1\mu}\delta_{2\nu} \nonumber \\ &
    +\left. \sum_{\bf{R}_n} e^{-i\kk\cdot\bf{R}_n} G^{\alpha\beta}_0(\bf{R}_n-\mathbf{d}_\mathrm{bas})\delta_{2\mu}\delta_{1\nu} \right] \nonumber,
\end{align}
where $\{\bf{R}_n\}$ represent the in-plane position vector of all the unit cells in the lattice. Additionally, $\{\alpha,\beta\}$ run over the 3 spatial degrees of freedom, and $\{\mu,\nu\}$ over the 2 lattice sites, such that $ \mathbf{M}_\kk$ is a 
$6\times6$ matrix with eigenvalues $\omega_\kk-i\frac{\gamma_\kk}{2}$, from which we obtain the photonic band structure and the radiative decay of the Bloch modes. The lattice sums of the Green's tensor are performed using the Ewald method~\cite{Ewald1921,Linton2010}.

Figure~\ref{fig:subwavelenth}(b,e) shows the band structure of a subwavelength isotropic honeycomb lattice ($d_\mathrm{inter}=d_\mathrm{intra}=0.15\lambda_a$) with the decay rate in color code, for modes where the dipole moments lie out of the plane (b) and in the plane (e), and for the path in the Brillouin zone sketched in the inset. In both cases we see Dirac points at K and K'~\cite{perczel17a}, as well as the characteristic features (polariton-like bends and divergences) when the modes interact with the light line. While modes within the light cone (gray area) couple to the free-space photonic continuum and hence present a non-zero decay rate, the Dirac cones appear outside the light cone where modes are confined to the array and thus subradiant.

Next we introduce uniaxial anisotropy in the metasurface, by allowing $\beta\neq 1$.  By calculating the band structures, we find two critical values of the anisotropy parameters where generalised Dirac dispersions emerge. First, for $\beta=0.85$ a semi-Dirac cone emerges for the out-of-plane modes. This is shown in panels (c,d), where we show the bands along the $\Gamma M$ direction displaying a Dirac crossing with linear dispersion at $M$ (panel c), while along the $K'MK$ direction the dispersion away from the degeneracy is quadratic (panel d). On the other hand, for $\beta=1.15$, two of the in-plane modes intersect forming a tilted Dirac cone (f,g). This occurs along the $\Gamma M$ line but away from any high symmetry points, as shown in (panel f). The dispersion in the perpendicular direction is plotted in (panel g), showing also the linear dispersion away from the degeneracy. 

Overall, these calculations show that both semi-Dirac and tilted Dirac energy dispersion can emerge in such subwavelength arrays. Then, as explained in Refs.~\cite{Masson2020,Patti2021,Brechtelsbauer2020}, in order to probe the physics explored along this manuscript one can place additional impurity atoms (as depicted in Fig.~\ref{fig:subwavelenth}(a)) with their optical transitions $\omega_0$ tuned to the spectral region where these non-trivial band-crossings occur, e.g., using Raman-assisted processes.

\section{Conclusion \& Outlook~\label{sec:conclu}}

Summing up, we have characterized the spontaneous emission properties of emitters spectrally tuned to several types of Dirac points using discrete photonic lattice models. For the case of anisotropic Dirac energy dispersions, we have shown how the shape of the emergent qubit-photon bound states and photon-mediated interactions can be modified without altering the asymptotic decay of the isotropic Dirac situation. This tuning can be done until the critical point (semi-Dirac case) where a directional localized quasi-bound state appears with a longer-ranged spatial decay. Besides, we have also explored the dynamical and radiation features of the case of tilted Dirac points, including a model that features a perfect critical tilting (type III Dirac point). There, we observe how the Dirac-Type I-III-II transition manifests in reversible decay dynamics and directional emission patterns. Finally, we demonstrate that these energy dispersions can be obtained using subwavelength atomic arrays despite the intrinsically long-range nature of their interaction when compared to the simplified models. Thus, this platform appears as a very suitable one to probe these phenomena using additional impurity atoms.

An interesting outlook of this work is to consider other types of two-dimensional band-touchings, such as purely parabolic ones that appear in bilayer systems~\cite{mele10a}, to study the interplay of these phenomena with different boundary conditions, or to harness non-local light-matter couplings~\cite{kockum18a,FriskKockum2021,Kannan2020,Wang2021a,Gonzalez-Tudela2019b} to be further engineer the quantum emitter dynamics.

\begin{acknowledgements}
 AGT acknowledges support from   CSIC Research   Platform   on   Quantum   Technologies   PTI-001  and  from  Spanish  project  PGC2018-094792-B-100(MCIU/AEI/FEDER, EU). AGT also acknowledges initial discussions of this project with U. Zabaleta. P.A.H. acknowledges funding from Funda\c c\~ao para a Ci\^encia e a Tecnologia and Instituto de Telecomunica\c c\~oes under project UIDB/50008/2020 and the CEEC Individual program from Funda\c c\~ao para a Ci\^encia e a Tecnologia with reference CEECIND/02947/2020. M.B.P. acknowledges support from the Spanish Ministerio de Ciencia e Innovación (PID2019-109905GA-C2) and from Eusko Jaurlaritza (IT1164-19 and KK-2019/00101).

\end{acknowledgements}

\bibliographystyle{apsrev4-1}
\bibliography{references}

\end{document}